\documentclass[useAMS,usenatbib]{mn2e}
\usepackage{amsmath}
\usepackage{graphicx}
\usepackage{epstopdf}
\usepackage{amssymb,latexsym}
\usepackage{multicol}
\usepackage{color}
\usepackage{pdflscape}

\begin{document}

\title[Open cluster Cepheids in {\sl Gaia} EDR3]{Galactic open cluster Cepheids -- a census based on {\sl Gaia} EDR3}

\author[Xiaoyue Zhou and Xiaodian Chen] {Xiaoyue Zhou$^{1}$ and Xiaodian
  Chen$^{1,2,3}$\thanks{E-mail: chenxiaodian@nao.cas.cn}\\
$^{1}$Department of Astronomy, China West Normal University, Nanchong, 637009, China\\
$^{2}$CAS Key Laboratory of Optical Astronomy, National Astronomical Observatories,
Chinese Academy of Sciences, Beijing 100101, China\\
$^{3}$School of Astronomy and Space Science, University of the Chinese Academy of Sciences, Beijing 101408, China}

\date{xxx}

\pagerange{\pageref{firstpage}--\pageref{lastpage}} \pubyear{2021}
\label{firstpage}

\maketitle

\begin{abstract}
In the {\sl Gaia} era, the membership analysis and parameter determination of open clusters (OCs) are more accurate. We performed a census of OC's classical Cepheids based on {\sl Gaia} Early Data Release 3 (EDR3) and obtained a sample of 33 OC Cepheids fulfilling the constraints of the spatial position, proper motion, parallax and evolution state. 13 of 33 OC Cepheids are newly discovered. Among them, CM Sct is the first first-crossing Cepheids with direct evidence of evolution. DP Vel is likely a fourth- or fifth-crossing Cepheids. Based on independent distances from OCs, $W_1$-band period-luminosity relation of Cepheids is determined with a 3.5\% accuracy: $\langle M_{W1} \rangle = -(3.274 \pm 0.090) \log P-(-2.567 \pm 0.080)$. The $\sl Gaia$-band period-Wesenheit relation agrees well with \citet{2019A&A...625A..14R}. A direct period-age relation for fundamental Cepheids are also determined based on OC's age, that is $\log t  = -(0.638 \pm 0.063) \log P+(8.569 \pm 0.057)$.
\end{abstract}


\section{Introduction}
Classical Cepheids are yellow giants or supergiants, intermediate-mass stars. Their temperatures are between 6000 and 8000 K, their luminosities are as high as 10$^{5}$ times the sun, and their periods are about 1 to 100 days. As one of the variable stars, Cepheid is famous for its period-luminosity relationship (PLR), also known as the Leavitt law. About a century ago, \citet{Leavitt12} first discovered the relationship between periods and luminosities base on 25 Cepheids in the Small Magellanic Cloud. Subsequently, astronomers strived to optimize the zero point of Cepheid's PLRs and use PLRs to determine distances of many galaxies (e.g., \citet{Freedman01, Sandage06}). At present, Cepheid's PLR has become an important distance ladder for establishing the cosmic distance scale \citep{Feast99}, and an important method for accurately measuring the Hubble constant \citep{Freedman11, Riess11,2021ApJ...908L...6R}. At the same time, Cepheid is an ideal tracer for studying the detailed disc structure of the Milky Way \citep{2019NatAs...3..320C,2019Sci...365..478S}. Recently, a number of surveys are dedicated to searching Cepheids in the Milky Way, including All-Sky Automated Survey for Supernovae \citep[ASAS-SN;][]{2018MNRAS.477.3145J}, the Asteroid Terrestrial-impact Last Alert System \citep[ATLAS;][]{2018AJ....156..241H}, the Wide-field Infrared Survey Explorer ({\sl WISE}) catalogue of periodic variable stars \citep{2018ApJS..237...28C}, the variable catalogue of {\sl Gaia} DR2 \citep{2019A&A...622A..60C}, the Cepheid catalogue of Optical Gravitational Lensing
Experiment \citep[OGLE;][]{2018AcA....68..315U} and the variable catalogue of Zwicky Transient Facility \citep[ZTF;][]{2020ApJS..249...18C}.

Open clusters (OCs) are groups of hundreds to thousands of stars with similar age, metallicity, and distance. Compared with globular clusters, they are younger and distributed closer to the galactic disc. The typical age of classical Cepheids is around 10 million years and they exist in young OCs of similar age. Since \citet{Irwin55} first discovered Cepheids in Galactic OCs, namely S Nor in NGC 6087 and U Sgr in M25, a lot of work has been devoted to this field, to finding new OC Cepheids, such as \citet{Feast57, van den Bergh57, Efremov64, Tsarevsky66, Turner86, Turner10, Turner93, Baumgardt00, An07, Majaess08}. The OC and Cepheid pairs have similar age, distance, proper motion, and metallicity. The properties of Cepheid can be inferred from their host cluster since properties of OC are easier to determine. By fitting the main-sequence of OC members, the independent distance can be obtained as the external calibrator of Cepheid's PLR \citep{An07, Turner10, Anderson13, 2017MNRAS.464.1119C, 2020A&A...643A.115B}. Cepheid with a longer period was found to have greater luminosity and mass, so it should evolve faster and younger. A series of studies (e.g., \citet{Efremov78, Magnier97, Efremov98, Grebel98, Efremov03, Bono05, Inno15, Senchyna15, Anderson16, De Somma20}) in terms of theory and observation are focused on Cepheid's period-age relation (PAR). The age of OC is used as the only observational evidence to limit Cepheid's PAR. These all mean the importance of a larger sample size of OC Cepheids.

In the last decade, \citet{Anderson13} took an 8-dimensional all-sky census to identify bona-fide cluster Cepheids, based on published catalogues or literature. \citet{2015MNRAS.446.1268C,2017MNRAS.464.1119C} used near-infrared (NIR) data to search for new OC Cepheids and enhance the probability of the known OC Cepheids. Afterwards, \citet{Clark15, Lohr18, Negueruela20, Alonso-Santiago20} performed a detailed analysis of several individual star clusters based on the spectral information, and discovered or confirmed 6 OC Cepheids. Currently, the number of Cepheids in the Milky Way OCs is still relatively small, only about 30. The robust membership also needs better astrometric data to confirm. In the {\sl Gaia} Early Data Release 3 (EDR3) era, the accuracy of astrometry has been greatly improved \citep{2020arXiv201201533G}. Accurate proper motions and parallaxes can be used to better find and confirm the membership of Cepheids and OCs. At the same time, based on more accuracy photometry of {\sl Gaia} EDR3, we can obtain better distance, extinction and age from clusters' isochrone fitting. In this work, we aim to conduct a census of OC Cepheids based on {\sl Gaia} EDR3 and determine the PLR and PAR only depending on OC Cepheids.

In Section 2, we introduce the adopted data and methods. We establish a new catalogue of OC Cepheids in Section 3, and also analyze the membership the error budget in this section. The PLR and PAR are discussed in Section 4. The whole work is summarized and prospected in Section 5.

\section{Data and method}
This section introduces the data adopted and how we establish the membership between Cepheids and OCs based on the spatial position, proper motion, age, parallax, and colour-magnitude diagram (CMD). We also explain how we determine the parameters of OCs based on isochrone fitting method and how we make an error budget.

\subsection{Data}
For the cluster sample, we used the cluster catalogue and cluster members catalogue from \citet{Cantat-GaudinAnders20}. In that work, they did a more thorough census of cluster members based on {\sl Gaia} DR2 \citep{Gaia Collaboration18} for about 2000 clusters and estimated the average proper motion and parallax for each cluster. The OC parameters such as age, extinction and distance are also determined in that work based on the artificial neural network method. These parameters were adopted to select OC Cepheids. For consistency, the astrometric and photometry data from {\sl Gaia} DR2 were adopted for searching OC Cepheids, while {\sl Gaia} EDR3 data were adopted for subsequent analysis. The photometric data is mainly based on {\sl Gaia} {\sl G}, {\sl G$_{\rm BP}$}, and {\sl G$_{\rm RP}$}. We also combined Pan-STARRS1 \citep{2016arXiv161205560C} and Two Micron All-Sky Survey \citep[2MASS]{2006AJ....131.1163S} photometry to reduce the uncertain of extinction. Pan-STARRS1 is a 3$\pi$ survey in five bands with a typical photometry accuracy better than 0.01 mag. 2MASS $J, K_{\rm S}$ were used for OCs in the southern sky which is not covered by Pan-STARRS1.

For the classical Cepheid, our samples include Cepheids collected in the General Catalogue of Variable Stars (GCVS) and the latest catalogues of variable stars, including ASAS-SN, ATLAS, {\sl WISE}, ZTF and OGLE, totaling about 3300. Considering that Cepheids are brighter than most of OC members, the sample of Cepheids is much more complete than that of OCs.

\subsection{OC Cepheid selection}
The OC Cepheid pairs are chosen from $\sim$2000 OCs and $\sim$3300 Cepheids. Faced with such a large sample, we set some preliminary selection criteria to reduce the workload. The spatial position is the first constraint for selecting OC Cepheid candidates. Here, we use two criteria: the candidates must be located within $3r_{50}$ or one degree from the cluster's centre. $r_{50}$ denotes the radius containing half the OC members, and we adopted the value estimated by \citet{Cantat-GaudinAnders20}. In most cases, the second criterion is looser than the first one, and there is little chance that an OC Cepheid can be located one degree away from the cluster.

Proper motion is a very powerful tool for determining the members of the cluster. It can separate most of the foreground stars and a few background stars from cluster members since cluster members have a similar direction of motion. We compared the proper motions of Cepheids with the average proper motion of the cluster, and chose OC Cepheid candidates with the proper-motion difference less than 3$\sigma_{\rm PM}$. $\sigma_{\rm PM}$ is the combined proper-motion uncertainty of OC and Cepheid.

Age is regarded as another independent constraint since Cepheid is very young with the age around $\log (t \mbox{ yr}^{-1}) =8.0$. We only choose OCs with age less than 8.6 to search for possible OC Cepheids. Given the age determination is more accurate for an old OC ($\log (t \mbox{ yr}^{-1}) >8.5$) than a young OC ($\log (t \mbox{ yr}^{-1}) <7.5$), we did not set a lower limit for the age criterion.

Parallax is the last criteria. {\sl Gaia} parallax is not accurate enough for Cepheids farther than $4-5$ kpc, but it can provide strong constraint for nearby Cepheids. We select OC Cepheid candidates by the parallax difference less than three times the parallax uncertainty $\sigma_{\rm plx}$. After preliminary selection from the spatial position, proper motion, age, and parallax, we got 55 OC Cepheid candidates that satisfy all criteria. We also obtained 290 candidates fulfill 3 of 4 criteria, but with proper-motion difference larger than 3$\sigma_{\rm PM}$ or position difference between $3r_{50}$ and 1 degree. For these 345 candidates, we plotted their CMDs to  remove non-cluster Cepheids further.

\begin{figure*}
\centering
  \begin{minipage}{160mm}
  \includegraphics[width=160mm]{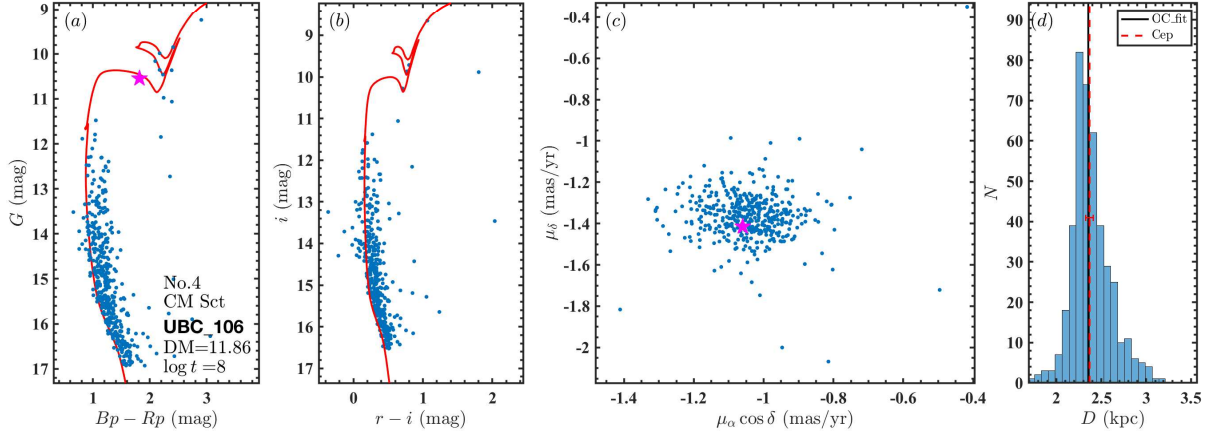}
\caption{The diagnostic diagram of OC Cepheid in UBC 106. Blue dots and red star symbol represent the OC's members and Cepheid, respectively. (a) Based on {\sl Gaia} CMD, the age, distance and extinction of the OC are obtained by the isochrone fitting method. The red line is the Padova isochrone. (b) Pan-STARRS1 CMD. (c) The proper-motion distributions of the OC and Cepheid. (d) Comparison of distances determined by $\sl Gaia$ EDR3 parallax and isochrone fitting (black solid line). The blue histogram shows the distribution of parallax distance of OC members, while the red dashed line and represents Cepheid's parallax distance. \label{c2f1.fig}}
  \end{minipage}
\end{figure*}

Majority of classical Cepheids are in the stage of helium core burning, so the genuine OC Cepheids are located close to the blue loop of their host cluster in CMD. We used the isochrone (see Section 2.3) to fit the main-sequence of OCs, and compared it with the Cepheids' loci (Figure \ref{c2f1.fig}a). By adopting this criterion, many background Cepheids or unlikely Cepheids (rotational stars) with magnitude fainter than the OC's turn-off point were excluded. A few Cepheids with redder or bluer colour are also excluded. Table \ref{c2table3} in the Appendix A lists 30 Cepheids that fulfill all criteria but are not genuine OC members. Of which, 5 Cepheids are close to the cluster's blue loop, and we assume them as possible OC Cepheids. Detailed studies are needed to make sure whether they are OC Cepheids in specific conditions or not. After selection, 33 OC Cepheid pairs were remaining. The CMDs, the diagram of proper motion and parallax comparison of two sample OC Cepheids are shown in Fig. \ref{c2f1.fig} and \ref{c2f2.fig}. The similar figures for other OC Cepheids are added in the Appendix. Detailed explanations of these figures can be found in section 2.3 and section 3. Note that {\sl Gaia} EDR3 data are used in these figures instead of {\sl Gaia} DR2 data. Compared to {\sl Gaia} DR2, the accuracy of proper motions and parallaxes is improved by 30\%. Nevertheless, it does not affect the membership of our OC Cepheids.

\begin{figure*}
\centering
  \begin{minipage}{160mm}
  \includegraphics[width=160mm]{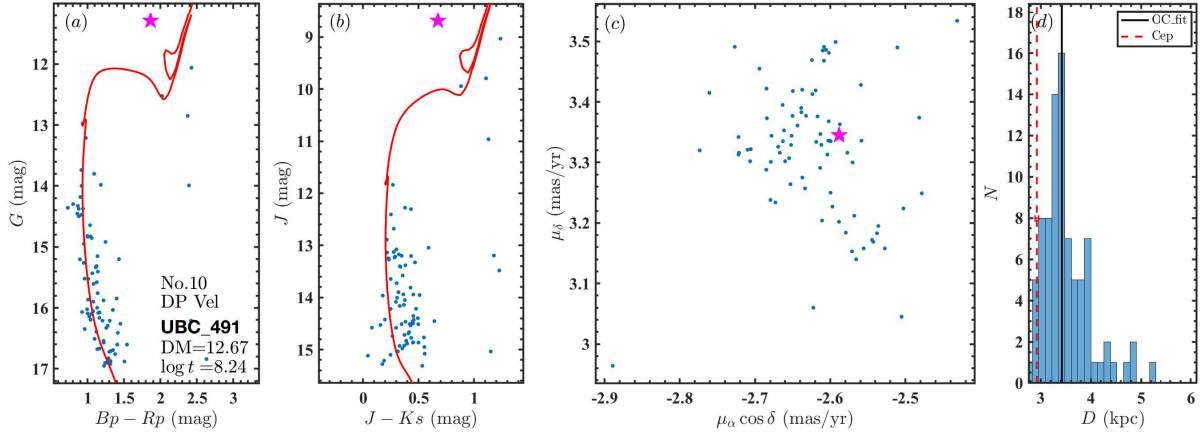}
\caption{Similar to Fig \ref{c2f1.fig} but for OC UBC 491. In panel (b), 2MASS CMD is adopted for OCs in the southern sky. \label{c2f2.fig}}
  \end{minipage}
\end{figure*}

\subsection{Isochrone fitting}
Isochrone fitting is an important method for measuring clusters' parameters, such as age, extinction and distance. We used the Padova isochrone with an age step of 0.01 dex to fit the OC's main-sequence. Since the metallicity of many OCs are unknown, we adopted a uniform solar metallicity for all OCs. This is a good approximation since our target OCs are within 4 kpc and close to the mid-plane. Age, distance and extinction from \citet{Cantat-GaudinAnders20} are used as the initial values, if they are given. The accuracy of these parameters, especially age, is not as good as that determined in the dedicated work \citep{2021arXiv210302275A,2021AJ....161..101S,2021arXiv210204303P}, so we manually arranged these parameters to fit each OC's main-sequence well with the isochrone.

During the fitting process, extinction is adjusted first to fit the zero-age main-sequence, which is close to the left edge of the main-sequence distribution (see Fig. \ref{c2f1.fig}(a)). We adopted the extinction law from \citet{2019ApJ...877..116W} with ${\sl Gaia}$ bands correction from \citet{2018A&A...616A..10G}. For the OC with lower extinction, the zero-age main-sequence is notable and the determination of extinction is relatively easier. While for several OCs with larger differential extinction, since zero-age main-sequence is obscure, we fit the main-sequence ridge line instead. The Pan-STARRS1 bands and 2MASS bands were used here to avoid underestimating extinction (see Fig. \ref{c2f1.fig}(b)) since bands with longer wavelengths are less sensitive to extinction. Most of our Cepheids are saturated in Pan-STARRS1 bands, so we do not use these CMDs to select OC Cepheids. The shape of isochrone in the pre main-sequence and post main-sequence is very sensitive to the age. Thus for OCs with more members in pre main-sequence stage and post main-sequence stage, the age accuracy is higher. Distance is mainly constrained by the lower main-sequence stars, and in some cases is related to the age determination. For OCs with no lower main-sequence or post main-sequence stars, age and distance are degenerate. That is increasing the age by 0.1 dex has the same effect as increasing the distance modulus of 0.25 mag. After isochrone fitting, we obtained the optimal extinction, age, and distance for each cluster, and list them in Table \ref{c2table1}. Cepheid shares the same parameters with its host OC. We compared the best-fitting distance with OC's and Cepheid's distances converted from the {\sl Gaia} parallax with a 17 $\mu$as offset correction (see Fig. \ref{c2f1.fig}(d)), and found them agreed well with each other. The typical uncertainty of Cepheid parallaxes is about 10\% at a distance of 2 kpc. We only used {\sl Gaia} parallax to provide a weak constraint on selecting OC Cepheids, and we use OC distances from main-sequence fitting to determine PLRs.

\subsection{Uncertainty of parameters}
Before using cluster parameters to study PLRs, we must know their uncertainties, which indicate the quality of these parameters. However, there is no standard method to estimate these uncertainties, especially for young and sparse OCs. Here, we try to estimate them by a uniform approach.

For extinction, the error is dominated by the dispersion of the zero-age main-sequence. We interpolated the isochrone and chose main-sequence stars with colour bluer than the isochrone ($(Bp-Rp)-(Bp-Rp)_{\rm iso}<0$). The error of the colour excess $\sigma_{E(Bp-Rp)}$ is the standard deviation of these stars' colour. If the number of this sample is less than 10\% of the total number of the cluster, we select 10\% stars with the smallest $(Bp-Rp)-(Bp-Rp)_{\rm iso}$ as the sample to estimate error. The error of colour excess was converted to extinction error $\sigma_{A_V}$ by the extinction law. For OCs whose age is around $10^8$ yr, the age of the OC is mainly constrained by the evolved stars. The age error was estimated by $0.3/\sqrt{N_P}$, where $N_P$ is the number of star in the post main-sequence phase. So for an OC with only one Cepheid and no other evolved stars, the age error is 0.3 dex. 0.3 dex is relatively large and suitable for the upper limit of age error. The distance modulus error is dominated by the lower main-sequence star, and we estimated it by $\sigma_{\rm{DM}}=0.75/\sqrt{N_{lms}}$ mag. $N_{lms}$ is the number of OC members with $M_G>1.8$. Since different OCs have different distances and extinctions, we use absolute magnitude $M_G$ to make this criterion applicable to most of our clusters. Berkeley 51 and FSR 0158 do not contain lower main-sequence stars in our sample, so we assumed uncertainty of 0.75 mag to their distance modulus.

The distance modulus errors for OCs with a significant main-sequence are around or better than 0.1 mag. The age errors are 0.1-0.3 dex while the colour excess errors are $\sigma_{E(Bp-Rp)} = 0.005-0.03$ mag. All these uncertainties are reasonable for an OC with the age around $10^8$ yr. We list these uncertainties in Table \ref{c2table1}.

\section{Result}
Based on {\sl Gaia} EDR3 data, a total of 33 OC Cepheids have been obtained, 13 of which are newly discovered. Table \ref{c2table1} lists the names of OC Cepheid pairs, the parameters and their errors of OCs, and constraints of the astrometric data. Based on the isochrone fitting, 31 of 33 Cepheids are in the blue-loop stage, while the other two clearly deviate from the blue-loop sequence. CM Sct in UBC 106 is during the first crossing of the Cepheid instability strip (see Fig. \ref{c2f1.fig}). DP Vel in UBC 491 is likely in the fourth- or fifth-crossing stage (see Fig. \ref{c2f2.fig}). This is reasonable, given that the blue-loop stage has the longest time scale for an evolved intermediate-mass star. 13 new OC Cepheids and one excluded OC Cepheids are discussed separately in this section. The diagnostic diagrams can be checked from Figures in the Appendix B.

\subsection{New OC Cepheids}

\subsubsection{UBC 106 and CM Sct}
UBC 106 and CM Sct are perfectly matched in terms of the spatial position (9.4 arcmin), proper motions ($0.2\sigma$ and $0.9\sigma$), and the parallax ($0.1\sigma$). The mean radial velocity of CM Sct is 40.8 km/s from \citet{1995IBVS.4148....1F}, which is within the range of the mean radial velocity of 6 members $41.63\pm 1.14$ km/s. All these ascertain that CM Sct is a genuine OC Cepheid. The OC's parameters from our isochrone fitting are ${\rm DM} = 11.86\pm0.05$ mag and $\log (t \mbox{ yr}^{-1}) = 8.00\pm0.11$, which is in good agreement with the reference value, ${\rm DM} = 11.86$ mag and $\log (t \mbox{ yr}^{-1}) = 8.2$ from \citet{Cantat-GaudinAnders20}. In CMD, CM Sct is clearly in the first crossing of the instability strip. Two possible first-crossing Cepheids Polaris and HD 344787, are discussed in \citet{2013ApJ...762L...8T,2021arXiv210108553R}. Compared to them, CM Sct's mass and age are strictly constrained by the host OC. Thus it can place constraints on the pulsation theory and the period variation of the first-crossing Cepheid. We also noticed four stars in the blue-loop stage are not pulsation stars, at least not pulsation stars with a moderate amplitude. High-quality time-domain photometry is needed in the future to investigate the pulsation and the evolution state of these stars.

\subsubsection{UBC 491 and DP Vel}
DP Vel is located at approximately 12 arcmin from the centre of UBC 491, and its proper motion is consistent with the cluster in $0.6\sigma$. The parallax of DP Vel is within the 0.8$\sigma$ range of OC members' parallaxes. The OC's parameters ${\rm DM} = 12.67\pm0.15$ mag and $\log (t \mbox{ yr}^{-1}) = 8.24\pm0.17$ from isochrone fitting are consistent with ${\rm DM} = 12.87$ mag and $\log (t \mbox{ yr}^{-1}) = 8.26$ from \citet{Cantat-GaudinAnders20}. DP Vel is found $\sim 0.7$ mag brighter than the blue-loop sequence in $G$-band, so it is likely a classical Cepheid in the asymptotic giant branch or the thermal pulsation stage. To confirm this, the age of UBC 491 should be better constrained, and the membership of DP Vel needs to be verified by radial velocity measurement.

\subsubsection{NGC 103 and NO Cas}
NO Cas is a first-overtone pulsator located at 8.5 arcmin from the centre of NGC 103. The proper motion of NO Cas is in good agreement with the mean proper motion of NGC 103 in $0.7\sigma$. From the CMD, we found that NO Cas is in the blue-loop stage. The derived distance modulus and age are ${\rm DM} = 12.60\pm0.07$ mag and $\log (t \mbox{ yr}^{-1}) = 8.21\pm0.21$, respectively. These match with the cluster parameters given by \citet{Cantat-GaudinAnders20}, ${\rm DM} = 12.5$ mag and $\log (t \mbox{ yr}^{-1}) = 8.01$. In addition, the difference between the parallax of NO Cas and NGC 103 is in 0.2$\sigma$. NO Cas meets all the criteria, and our results support it as an OC Cepheid.
\subsubsection{Kronberger 84 and J213533.70+533049.3, V733 Cyg}
J213533.70+533049.3 is found about 0.3 arcmin from the centre of Kronberger 84, while V733 Cyg is located at about 18.2 arcmin from the cluster centre. The difference between their respective proper motions and the mean proper motion of Kronberger 84 is 1.0$\sigma$ and 2.0$\sigma$. In the proper-motion diagram, V733 Cyg seems a little away from the cluster's mean proper motion, which is due to the small number of OC members. Both the two Cepheids are in the blue-loop stage, and their parallaxes agree well with the cluster's parallax in $0.2\sigma$. The result of isochrone fitting leads to a distance modulus of ${\rm DM} = 13.03\pm0.21$ mag and an age of $\log (t \mbox{ yr}^{-1}) = 8.16\pm0.21$. The values are consistent with \citet{Cantat-GaudinAnders20}. J213533.70+533049.3 and V733 Cyg meet all the criteria and are high-probability OC Cepheids.
\subsubsection{UBC 229 and V335 Pup}
V335 Pup is located at about 2.4 arcmin from the centre of UBC 229. Its proper motion is very close to the mean proper motion of UBC 229. According to the fit results, V335 Pup is in the blue-loop stage, and the parameters of UBC 229 are ${\rm DM} = 11.94\pm0.12$ mag and $\log (t \mbox{ yr}^{-1}) = 7.95\pm0.30$, They are agreed well with ${\rm DM} = 12.04$ mag and $\log (t \mbox{ yr}^{-1}) = 8.05$ from \citet{Cantat-GaudinAnders20}. V335 Pup's parallax is 0.4$\sigma$ from the mean parallax of cluster's members. V335 Pup satisfies all our criteria at the same time.

\begin{table*}
\tiny
 \begin{minipage}{200mm}
\caption{Open cluster Cepheid catalogue.\label{c2table1}}
\begin{tabular}{@{}lccclccccccc@{}}
  \hline
Cluster  & DM & $\log$ Age  & $A_V$ & Cepheid   &  $\Delta (\mu_\alpha\cos\delta)$  &  $\Delta \mu_\delta$ & $\Delta parallax$ & Seperation &$\log P$ & new(Y) &Num  \\
         & (mag)& [yr]   &(mag) &           &                  &      &     & (arcmin)  &  (d) &  & \\
  \hline

NGC 103       & $12.60\pm0.07$ & $8.21\pm0.21$ & $1.30\pm0.03$ & NO Cas                     &1.2$\sigma$  & 0.4$\sigma$   & 0.2$\sigma$ & 8.5  & 0.412 &Y &1  \\
NGC 6664      & $11.42\pm0.08$ & $8.11\pm0.12$ & $1.94\pm0.04$ & EV Sct                     &1.1$\sigma$  & 0.2$\sigma$   & 0.4$\sigma$ & 2.4  & 0.490 &N &2 \\ 
Kronberger 84 & $13.03\pm0.21$ & $8.16\pm0.21$ & $2.13\pm0.09$ & J213533.70+533049.3        &1.0$\sigma$  & 0.9$\sigma$   & 0.1$\sigma$ & 0.3  & 0.506 &Y &3 \\ 
UBC 106       & $11.86\pm0.05$ & $8.00\pm0.11$ & $2.11\pm0.06$ & CM Sct                     &0.2$\sigma$  & 0.9$\sigma$   & 0.1$\sigma$ & 9.4  & 0.593 &Y &4 \\ 
NGC 129       & $11.40\pm0.05$ & $7.96\pm0.17$ & $1.51\pm0.05$ & V379 Cas                   &0.4$\sigma$  & 0.7$\sigma$   & 0.0$\sigma$ & 44.9 & 0.634 &N &5 \\ 
Berkeley 58   & $12.40\pm0.09$ & $8.18\pm0.30$ & $1.83\pm0.02$ & CG Cas                     &4.1$\sigma$  & 2.0$\sigma$   & 1.1$\sigma$ & 5.5  & 0.640 &N &6 \\ 
NGC 7790      & $12.50\pm0.08$ & $8.13\pm0.17$ & $1.58\pm0.03$ & CE Cas B                   &0.9$\sigma$  & 1.9$\sigma$   & 0.6$\sigma$ & 2.2  & 0.651 &N &7 \\ 
Kronberger 84 & $13.03\pm0.21$ & $8.16\pm0.21$ & $2.13\pm0.09$ & V733 Cyg                   &1.7$\sigma$  & 2.1$\sigma$   & 0.2$\sigma$ & 18.2 & 0.659 &Y &  \\ 
UBC 229       & $11.94\pm0.12$ & $7.95\pm0.30$ & $0.62\pm0.02$ & V335 Pup                   &0.5$\sigma$  & 0.1$\sigma$   & 0.4$\sigma$ & 2.4  & 0.687 &Y &8 \\ 
NGC 7790      & $12.50\pm0.08$ & $8.13\pm0.17$ & $1.58\pm0.03$ & CF Cas                     &1.9$\sigma$  & 0.8$\sigma$   & 0.3$\sigma$ & 1.3  & 0.688 &N &  \\ 
NGC 7790      & $12.50\pm0.08$ & $8.13\pm0.17$ & $1.58\pm0.03$ & CE Cas A                   &0.1$\sigma$  & 0.8$\sigma$   & 0.3$\sigma$ & 2.1  & 0.711 &N &  \\ 
vdBergh 1     & $11.25\pm0.10$ & $8.09\pm0.30$ & $2.08\pm0.05$ & CV Mon                     &0.0$\sigma$  & 0.1$\sigma$   & 0.5$\sigma$ & 0.9  & 0.731 &N &9 \\ 
UBC 491       & $12.67\pm0.15$ & $8.24\pm0.17$ & $2.11\pm0.07$ & DP Vel                     &1.0$\sigma$  & 0.2$\sigma$   & 0.8$\sigma$ & 12.0 & 0.739 &Y &10 \\
NGC 5662      & $9.36 \pm0.05$ & $8.10\pm0.17$ & $0.78\pm0.01$ & V Cen                      &0.2$\sigma$  & 1.0$\sigma$   & 0.8$\sigma$ & 24.6 & 0.740 &N &11 \\
UBC 135       & $13.05\pm0.43$ & $7.98\pm0.30$ & $2.15\pm0.09$ & GI Cyg                     &1.6$\sigma$  & 2.2$\sigma$   & 0.3$\sigma$ & 3.6  & 0.762 &Y &12 \\
Berkeley 55   & $12.20\pm0.15$ & $8.10\pm0.12$ & $4.32\pm0.08$ & 2MASS J21165990+5145587    &0.7$\sigma$  & 0.3$\sigma$   & 0.4$\sigma$ & 0.7  & 0.767 &N &13 \\
Ruprecht 79   & $13.00\pm0.12$ & $8.07\pm0.12$ & $1.98\pm0.05$ & CS Vel                     &0.6$\sigma$  & 0.3$\sigma$   & 1.0$\sigma$ & 2.2  & 0.771 &N &14 \\
UBC 290	      & $10.98\pm0.08$ & $8.08\pm0.21$ & $0.65\pm0.01$ & X Cru                      &0.7$\sigma$  & 0.3$\sigma$   & 1.2$\sigma$ & 18.9 & 0.794 &Y &15 \\
NGC 6649      & $11.44\pm0.05$ & $7.93\pm0.12$ & $3.90\pm0.07$ & V367 Sct                   &0.4$\sigma$  & 0.2$\sigma$   & 0.5$\sigma$ & 2.8  & 0.799 &N &16 \\
UBC 129       & $10.31\pm0.08$ & $8.07\pm0.30$ & $1.33\pm0.03$ & X Vul                      &3.7$\sigma$  & 2.3$\sigma$   & 0.1$\sigma$ & 18.3 & 0.801 &Y &17 \\
IC 4725       & $9.02 \pm0.04$ & $8.05\pm0.21$ & $0.92\pm0.04$ & U Sgr                      &0.0$\sigma$  & 0.3$\sigma$   & 0.7$\sigma$ & 2.1  & 0.829 &N &18 \\
Ruprecht 97   & $12.95\pm0.53$ & $8.13\pm0.21$ & $1.61\pm0.01$ & SV Cru                     &0.7$\sigma$  & 10.9$\sigma$  & 0.3$\sigma$ & 4.7  & 0.845 &Y &19 \\
FSR 0951      & $11.22\pm0.07$ & $7.97\pm0.30$ & $1.05\pm0.02$ & RS Ori                     &0.7$\sigma$  & 0.4$\sigma$   & 0.3$\sigma$ & 2.0  & 0.879 &N &20 \\
NGC 129       & $11.40\pm0.05$ & $7.96\pm0.17$ & $1.51\pm0.05$ & DL Cas                     &0.3$\sigma$  & 3.4$\sigma$   & 1.6$\sigma$ & 3.4  & 0.903 &N &   \\
UBC 231       & $12.31\pm0.17$ & $7.87\pm0.30$ & $0.98\pm0.06$ & WX Pup                     &1.6$\sigma$  & 2.4$\sigma$   & 0.6$\sigma$ & 25.9 & 0.951 &Y &21 \\
NGC 6087      & $9.85 \pm0.06$ & $7.90\pm0.14$ & $0.55\pm0.01$ & S Nor                      &1.0$\sigma$  & 0.8$\sigma$   & 0.5$\sigma$ & 1.0  & 0.989 &N &22 \\
Berkeley 51   & $13.60\pm0.75$ & $7.98\pm0.11$ & $5.00\pm0.09$ & 2MASS J20115118+3424472    &1.0$\sigma$  & 0.5$\sigma$   & 0.9$\sigma$ & 0.8  & 0.993 &N &23 \\
Lynga 6       & $11.72\pm0.34$ & $7.94\pm0.21$ & $3.51\pm0.06$ & TW Nor                     &2.0$\sigma$  & 0.6$\sigma$   & 0.5$\sigma$ & 0.6  & 1.033 &N &24 \\
UBC 375       & $11.08\pm0.21$ & $8.03\pm0.30$ & $2.94\pm0.01$ & V438 Cyg                   &1.4$\sigma$  & 7.3$\sigma$   & 0.5$\sigma$ & 7.0  & 1.050 &Y &25 \\
NGC 6067      & $11.37\pm0.03$ & $8.00\pm0.08$ & $0.87\pm0.04$ & V340 Nor                   &1.3$\sigma$  & 0.5$\sigma$   & 0.5$\sigma$ & 1.0  & 1.053 &N &26 \\
FSR 0158      & $13.64\pm0.75$ & $7.81\pm0.25$ & $4.61\pm0.03$ & GQ Vul                     &0.8$\sigma$  & 0.2$\sigma$   & 0.1$\sigma$ & 2.4  & 1.102 &Y &27 \\
BH 222        & $13.65\pm0.21$ & $7.73\pm0.08$ & $6.00\pm0.51$ & 2MASS J17184740-3817292    &1.1$\sigma$  & 1.0$\sigma$   & 0.2$\sigma$ & 0.2  & 1.367 &N &28 \\
UBC 130       & $11.72\pm0.11$ & $7.47\pm0.17$ & $1.82\pm0.04$ & SV Vul                     &0.5$\sigma$  & 0.4$\sigma$   & 0.5$\sigma$ & 9.7  & 1.652 &N &29 \\

\hline
 \end{tabular}
\end{minipage}
\end{table*}

\subsubsection{UBC 135 and GI Cyg}
GI Cyg is found at about 3.6 arcmin from the centre of UBC 135. The proper motion of GI Cyg is $(\mu_{\alpha\cos\delta,{\rm Cep}},\mu_{\delta,{\rm Cep}}) = (-3.39 \pm 0.04, -6.25\pm 0.05)$ mas yr$^{-1}$, which is 2.0$\sigma$ away from the mean proper motion of UBC 135, $(\mu_{\alpha\cos\delta,{\rm cl}},\mu_{\delta,{\rm cl}}) = (-3.52 \pm 0.07, -6.44\pm 0.07)$ mas yr$^{-1}$. Our estimated age $\log (t \mbox{ yr}^{-1}) = 7.98\pm0.30$ is 0.6 dex older than that from \citet{Cantat-GaudinAnders20}. Our age determination is only based the Cepheid, so an 0.3 dex age uncertainty is assumed.

\subsubsection{UBC 290 and X Cru}
X Cru is found at about 18.9 arcmin (2.1$r_{50}$) from the centre of UBC 290. X Cru agrees well with its host cluster in the proper motion ($0.4\sigma$) and parallax ($1.2\sigma$). Based on two evolved stars, our estimated age $\log (t \mbox{ yr}^{-1}) = 8.08\pm0.21$ is only 0.2 dex younger than that from \citet{Cantat-GaudinAnders20}.

\subsubsection{UBC 129 and X Vul}
X Vul is found at about 18.2 arcmin (1.5$r_{50}$) from the centre of UBC 129. X Vul's parallax agrees well with its host cluster in $0.1\sigma$, while its proper motion is at the edge of the cluster's distribution ($3\sigma$). Based on the location of X Vul in CMD, our estimated age $\log (t \mbox{ yr}^{-1}) = 8.07\pm0.3$ is 0.35 dex older than that from \citet{Cantat-GaudinAnders20}.

\subsubsection{Ruprecht 97 and SV Cru}
SV Cru is at about 4.7 arcmin from the centre of Ruprecht 97. SV Cru satisfied our criteria except for the proper motion in declination direction. The proper motion of SV Cru, $(\mu_{\alpha\cos\delta,{\rm Cep}},\mu_{\delta,{\rm Cep}}) = (-6.43 \pm 0.04, 1.31\pm 0.05)$ mas yr$^{-1}$, deviates 10.9$\sigma$ in declination direction from the proper motion of Ruprecht 97, $(\mu_{\alpha\cos\delta,{\rm cl}},\mu_{\delta,{\rm cl}}) = (-6.49 \pm 0.07, 0.44\pm 0.07)$. The mean proper motion is estimated by 23 OC members and the deviation of proper motion may be caused by the strict selection of OC members. Considering the good matching of other constraints, we regard SV Cru as an OC Cepheid. The distance and age of Ruprecht 97 are ${\rm DM} = 12.95 \pm 0.53$ mag and $\log (t \mbox{ yr}^{-1}) = 8.13 \pm 0.21$, which are consistent with \citet{Cantat-GaudinAnders20}.

\subsubsection{UBC 231 and WX Pup}
WX Pup is located at about 25.9 arcmin from the centre of UBC 231. Its proper motion is 2.2$\sigma$ from the mean proper motion of UBC 231. The parameters of UBC 231 ${\rm DM} = 12.31\pm0.17$ mag and $\log (t \mbox{ yr}^{-1}) = 7.87\pm0.30$, are in agreement with the reference values ${\rm DM} = 12.31$ mag and $\log (t \mbox{ yr}^{-1}) = 8.07$, given by \citet{Cantat-GaudinAnders20}.
\subsubsection{UBC 375 and V438 Cyg}
V438 Cyg is located at approximately 7.0 arcmin from the centre of UBC 375. V438 Cyg satisfies the constraints of the spatial position and parallax. The proper motion of V438 Cyg is $(\mu_{\alpha\cos\delta,{\rm Cep}},\mu_{\delta,{\rm Cep}}) = (-3.28 \pm 0.07, -4.27\pm 0.07)$ mas yr$^{-1}$, which is 6.4$\sigma$ away from the mean proper motion of UBC 375 $(\mu_{\alpha\cos\delta,{\rm cl}},\mu_{\delta,{\rm cl}}) = (-3.42 \pm 0.07, -5.22 \pm 0.11)$ mas yr$^{-1}$. The determined OC's parameters are ${\rm DM} = 11.08\pm0.21$ mag and $\log (t \mbox{ yr}^{-1}) = 8.03\pm0.30$ which are consistent with \citet{Cantat-GaudinAnders20}. Since only 32 OC members are adopted to estimate the mean proper motion, the deviation of proper motion may be attributed to the strict selection of OC members. Considering the good matching of other constraints, we still regard V438 Cyg as an OC Cepheid.

\subsubsection{FSR 0158 and GQ Vul}
GQ Vul is located at approximately 2.4 arcmin from the centre of FSR 0158. GQ Vul agrees well with FSR 0158 in the proper motion ($0.6\sigma$) and parallax ($0.1\sigma$). This new OC Cepheid is $\sim5.3$ kpc from us and almost at the detection limit of our method based on $\sl Gaia's$ proper motions and parallaxes. The far away OC Cepheids in the Milky Way's disk are helpful to testify the probable bias in Cepheid's PAR.

\subsection{Issue about the known OC Cepheid}
In this subsection, we discuss known OC Cepheids not included in our catalog. We compared with \citet{Turner10} who established an OC Cepheid catalogue based on optical data and found that 11 of the 24 OC Cepheids in that work are in our list. 9 of the other 13 Cepheids are in small clusters or star associations and they are not included in Cantat-Gaudin's membership analysis catalog. 3 OC Cepheids have proper-motion differences larger than $10\sigma$. The remaining OC Cepheid is BB Sgr in Collinder 394.

\subsubsection{Collinder 394 and BB Sgr}
BB Sgr was treat as an OC Cepheid for a long time from \citet{1985AJ.....90.1231T}. However, based on {\sl Gaia}, the membership is questionable. The proper motion of BB Sgr is $(\mu_{\alpha\cos\delta,{\rm Cep}},\mu_{\delta,{\rm Cep}}) = (0.31 \pm 0.10, -5.14\pm 0.09)$ mas yr$^{-1}$, which is 6.6$\sigma$ away from the mean proper motion of Collinder 394 $(\mu_{\alpha\cos\delta,{\rm cl}},\mu_{\delta,{\rm cl}}) = (-1.46 \pm 0.18, -5.88 \pm 0.17)$ mas yr$^{-1}$. The parallax difference is $2\sigma$ and very larger for a star at a distance of 0.8 kpc. The mean radial velocity of BB Sgr is 4.9 km/s from \citet{1995IBVS.4148....1F}, which is in $1.6\sigma$ range of the mean radial velocity of 4 members $19.4\pm 9.1$ km/s.

\section{Discussion}
If an OC Cepheid's membership is ascertained, we can assume Cepheid shares the same age and distance with its host OC. These age and distance are independent measurements that can be used to determine the PLR and PAR of classical Cepheids. In this section, we establish the accurate PLR and PAR of Cepheids and compare them with the previous work.

\subsection{Period-Luminosity Relation}
Cepheid PLR is an important tool to determine the Hubble constant and study the Milky Way's structure. The $1\sigma$ scatter of infrared PLRs or period-Wesenheit relations (PWRs) of classical Cepheid is about $3-5\%$. The improvement of the zero point accuracy is essential to reduce the systematic error of the Hubble constant and the Milky Way's parameters. Here, we adopted the OC distance determined by isochrone fitting rather than the average parallax as the distance of Cepheid, since the parallax's systematic error is still quite uncertain for the disc star \citep{2020arXiv201201742L, 2021AJ....161..176R}. The best PLRs are determined by 16 OC Cepheids with distance modulus errors less than 0.12 mag. The $W1$-band PLR and $\sl Gaia$-band PWR are:
\begin{equation}\label{c2equation1}
 \begin{aligned}
 \langle M_{W1} &\rangle = -(3.274 \pm 0.090) \log P-(-2.567 \pm 0.080),\\ 
                &\sigma=0.075 \\
 \end{aligned}
\end{equation}
\begin{equation}\label{c2equation2}
 \langle W_{G, Bp, Rp} \rangle = -3.32 \log P-(2.718 \pm 0.049), \sigma=0.161 \\
\end{equation}
The average $W1$-band magnitude of Cepheids are from ALLWISE \citep{2010AJ....140.1868W}, while the $\sl Gaia$-band mean magnitudes are from \citet{2019A&A...625A..14R}. 10 Cepheids have no $\sl Gaia$-band mean magnitude are not considered in determining the PWR. We adopted Wesenheit magnitude from \citet{2019A&A...625A..14R}, that is $W_{G, Bp, Rp}=\langle G \rangle-1.90(\langle Bp \rangle-\langle Rp \rangle)$. Absolute magnitudes and pulsation modes of our OC Cepheids are list in Table \ref{c2table2}.

\begin{figure}
\centering
  \includegraphics[width=80mm]{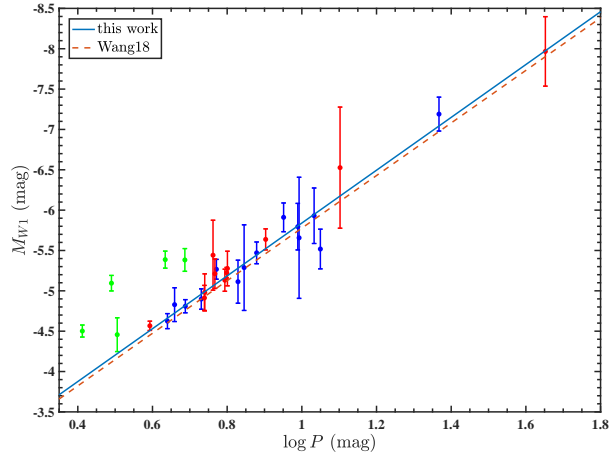}
\caption{PLR of OC Cepheids in the mid-infrared $W1$ band. The red, blue and green dots are new fundamental Cepheids in OCs, known fundamental Cepheids in OCs and overtone Cepheids in OCs. Our determined PLR based on 16 fundamental Cepheids with accurate distance is the blue line while the red dotted line is the $W1$-band PLR from \citet{Wang18}. \label{wpl.fig}}
\end{figure}

Fig. \ref{wpl.fig} shows the comparison of our $W1$-band PLR with that obtained by \citet{Wang18} based on hundreds of Milky Way's Cepheids. The dispersion of the two PLRs is comparable, which means that $W1$-band PLR can provide a distance with 3.5\% accuracy for Cepheids in the Milky Way. There is a small zero-point difference between the two PLRs at $P=10$ day, ${\rm PLR_{our} (W1) -PLR_{Wang18} (W1)}=-0.063$ mag. This difference is the result of different distance determination methods and the different zero points of $W1$-band photometry. The zero point of distance method could be tested by the future $\sl Gaia$ parallaxes with better constrained systematic errors for young stars. ALLWISE $W1$-band magnitude is the average magnitudes of data collected for a total of 2-3 days, which is not enough to cover the entire period of most Cepheids. Although NEOWISE covers the other 12 days for most Cepheids so far this year, the zero-point difference is significant between ALLWISE and NEOWISE for Cepheids brighter than 8 mag \citep{2014ApJ...792...30M}, which makes a simple correction based on NEOWISE for our OC Cepheids unreliable. The accurate determination of $W1$-band mean magnitude is important for Cepheids and other variables and deserves future work. The 5 overtone Cepheids are $0.61\pm0.16$ mag brighter than the fundamental Cepheids at the same period.

\begin{table*}
\footnotesize
 \begin{minipage}{120mm}
\caption{Cepheid's absolute magnitude.\label{c2table2}}
\begin{tabular}{@{}lcccccc@{}}
  \hline
Cepheid  &  $\log P$   & $M_{W1}$ & $\sigma(M_{W1})$  &  $M_{G, BP, RP}$  & $\sigma(M_{G, BP, RP})$  & Mode  \\
         &  [d]        &(mag)     &  mag              & mag             &       mag           &     \\
  \hline

NO Cas                 &  0.412& -4.502& 0.075& -3.758& 0.077&   o \\
EV Sct                 &  0.490& -5.095& 0.096& -4.995& 0.093&   o \\
J213533.70+533049.3    &  0.506& -4.457& 0.209&       &      &   o \\
CM Sct                 &  0.593& -4.567& 0.057& -4.940& 0.064&   f \\
V379 Cas               &  0.634& -5.386& 0.106& -5.621& 0.060&   o \\
CG Cas                 &  0.640& -4.625& 0.093& -4.635& 0.109&   f \\
CE Cas B               &  0.651&       &      & -4.818& 0.106&   f \\
V733 Cyg               &  0.659& -4.829& 0.209&       &      &   f \\
V335 Pup               &  0.687& -5.383& 0.141& -5.537& 0.118&   o \\
CF Cas                 &  0.688& -4.809& 0.081& -4.950& 0.095&   f \\
CE Cas A               &  0.711&       &      & -4.867& 0.107&   f \\
CV Mon                 &  0.731& -4.898& 0.127&       &      &   f \\
DP Vel                 &  0.739& -4.913& 0.155& -5.129& 0.172&   f \\
V Cen                  &  0.740& -4.981& 0.229& -5.109& 0.088&   f \\
GI Cyg                 &  0.762& -5.442& 0.433& -5.696& 0.438&   f \\
2MASS J21165990+5145587&  0.767& -5.212& 0.182&       &      &   f \\
CS Vel                 &  0.771& -5.267& 0.123& -5.436& 0.136&   f \\
X Cru                  &  0.794& -5.131& 0.137& -5.355& 0.084&   f \\
V367 Sct               &  0.799& -5.225& 0.065&       &      &   f(o) \\
X Vul                  &  0.801& -5.277& 0.214& -5.680& 0.085&   f \\
U Sgr                  &  0.829& -5.113& 0.267& -5.371& 0.054&   f \\
SV Cru                 &  0.845& -5.287& 0.531& -5.207& 0.536&   f \\
RS Ori                 &  0.879& -5.471& 0.135& -5.708& 0.124&   f \\
DL Cas                 &  0.903& -5.638& 0.130& -5.889& 0.072&   f \\
WX Pup                 &  0.951& -5.911& 0.178& -6.048& 0.179&   f \\
S Nor                  &  0.989& -5.795& 0.288& -5.974& 0.092&   f \\
2MASS J20115118+3424472&  0.993& -5.657& 0.750&       &      &   f \\
TW Nor                 &  1.033& -5.931& 0.344&       &      &   f \\
V438 Cyg               &  1.050& -5.518& 0.246&       &      &   f \\
V340 Nor               &  1.053&       &      & -6.126& 0.039&   f \\
GQ Vul                 &  1.102& -6.526& 0.750&       &      &   f \\
2MASS J17184740-3817292&  1.367& -7.190& 0.210&       &      &   f \\
SV Vul                 &  1.652& -7.967& 0.430& -8.463& 0.141&   f \\
\hline
 \end{tabular}
\end{minipage}
\end{table*}

\begin{figure}
\centering
  \includegraphics[width=80mm]{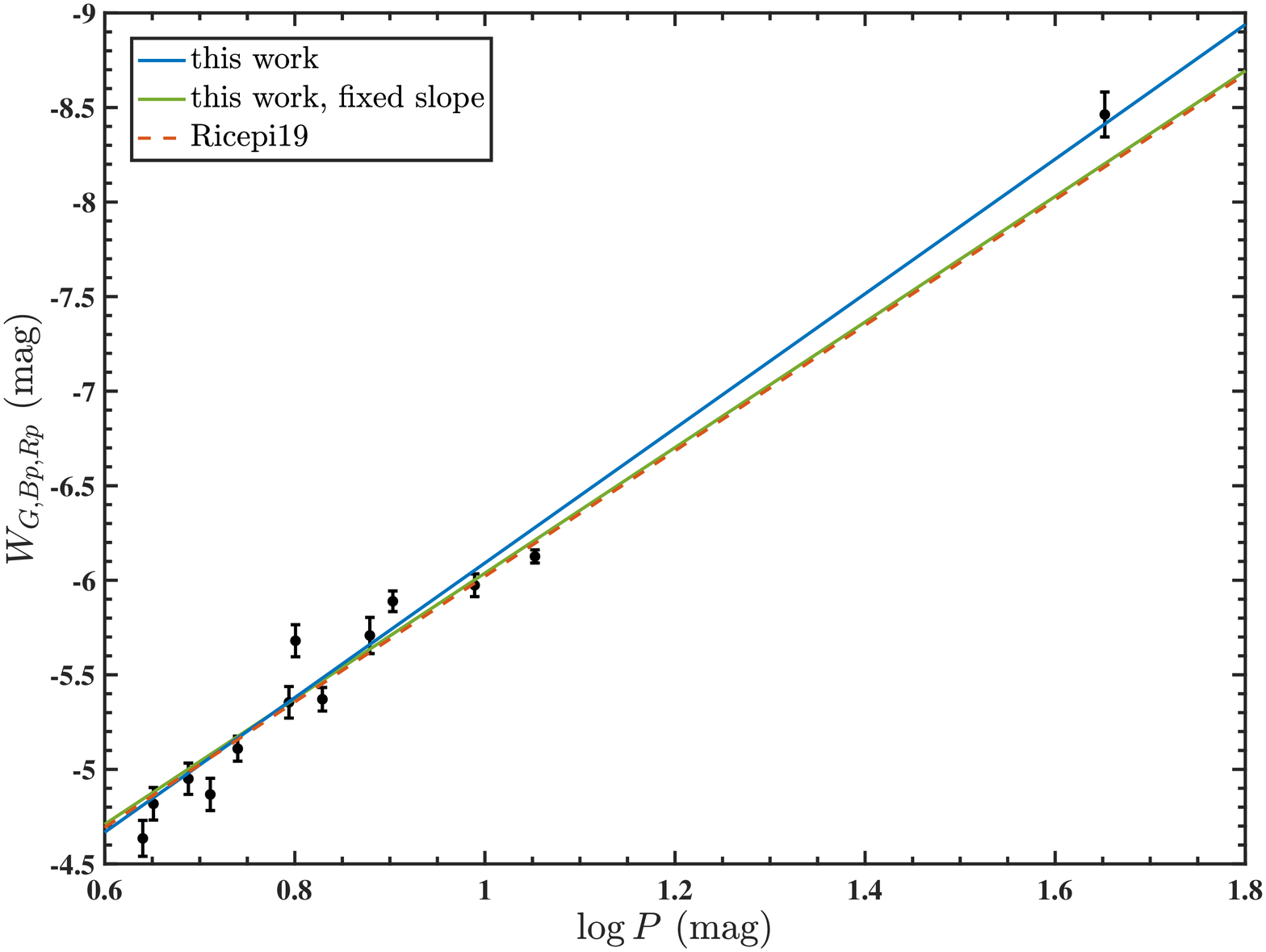}
\caption{Comparison of our PWR for fundamental Cepheids with that from \citet[red dashed line]{2019A&A...625A..14R}. The green line represents our PWR with a fixed slope. \label{gpl.fig}}
\end{figure}

 Compared to $W1$ band, the $Gaia$-band photometry has negligible zero point problem. Fig. \ref{gpl.fig} shows the comparison of $Gaia$-band PWR with \citet{2019A&A...625A..14R}. The slopes of two PWRs are different, and we guess the main reason is that our PWR is only based on 13 Cepheids. If we fix our slope to \citet{2019A&A...625A..14R}, the two PWRs agree well with each other, with only a $-0.017$ mag difference. We noted that the scatter of $Gaia$-band PWR is about twice of $W1$-band PLR. This larger scatter is caused by Gaia's wide band. As mentioned in \citet{2019ApJ...877..116W}, the variation of the extinction law in $G$-band is ten times that in $V$ band. The coefficient of $Gaia$-band PWR $1.90$ is not an approximate constant but depends on the amount of extinction and the intrinsic colour. An item $(\langle Bp \rangle-\langle Rp \rangle)^2$ added in the Wesenheit function may be helpful to reduce the scatter.

\subsection{Period-Age Relation}
\begin{figure}
\centering
  \includegraphics[width=80mm]{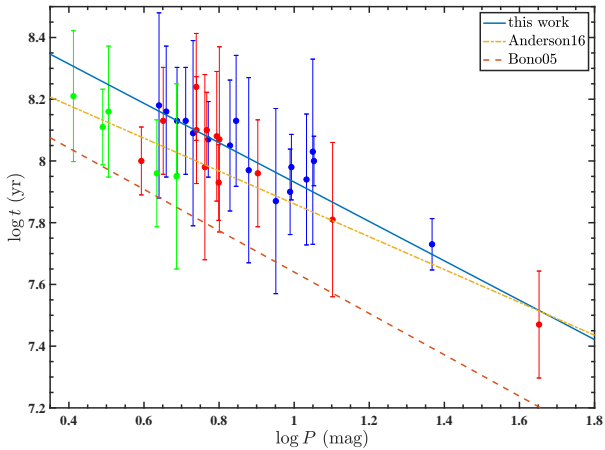}
\caption{Period-age relation of 33 OC Cepheids. The red, blue and green dots are new fundamental Cepheids in OCs, known fundamental Cepheids in OCs and overtone Cepheids in OCs, respectively. The blue line is the fitted PAR of fundamental Cepheids, which is compared with those from \citet[red dashed line]{Bono05} and \citet[yellow dash-dotted line]{Anderson16}. \label{pa.fig}}
\end{figure}

As a star in the blue-loop stage, the more massive Cepheid variable star is, the younger it is. Given Cepheids comply with PLR, they also comply with PAR. In previous studies, PAR was mainly obtained based on the stellar evolution theory with some observational constraints \citep{Bono05, Anderson16, De Somma20}. The direct age from OCs is important to constrain the zero point of PAR. However, it is not easy to establish a large sample of Cepheid with direct age because it requires a detailed analysis of OC and Cepheid pairs and requires a sufficiently accurate age determination of OCs. Here, we use 33 OC Cepheids to determine the PAR (see Fig. \ref{pa.fig}). As for the fundamental Cepheids, the equation of robust linear fit is $\log t  = -(0.638 \pm 0.063) \log P+(8.569 \pm 0.057), \sigma=0.073$. The $1\sigma$ scatter is about 17\% in age which is quite good for the age determination. Compared to fundamental Cepheids, the first overtone Cepheids are 0.14 dex (about 32\% in age) younger for a given period.

\citet{Bono05} supplied a theoretical calibration of the PAR by adopting homogeneous and detailed sets of nonlinear, convective pulsation models together with up-to-date evolutionary models. \citet{Anderson16} studied the PAR of Cepheids based on the stellar structure from rotating models. Compared with these works, we found that our PAR of fundamental Cepheids is systematically 0.3 dex older than \citet{Bono05}, but the slopes are consistent. Our PAR is more consistent with \citet{Anderson16}, although there is still a small deviation (0.15 dex) in the short-period end. This deviation could only be better studied with the discovery of short-period Cepheids ($\log P<0.6$) in relatively older OCs. Considering that we used a larger sample size of OC Cepheids and better ages of OCs, our PAR's slope and zero point are more accurate. By adopting PARs of both fundamental and first-overtone Cepheids, the Milky Way's age distribution can be better investigated, thus verifying the formation and evolution theory of the disc structure and spiral arms.

\section{Conclusion}
Based on {\sl Gaia} EDR3 data, we performed a census of OC Cepheids using a sample of 2000 OCs with parameters given by \citet{Cantat-GaudinAnders20} and a sample of 3300 Cepheids collected from the literature. To select OC Cepheids, we established five criteria: (1) the spatial positions of OC's centre and Cepheid are in one degree or $3r_{50}$; (2) the difference between Cepheid's proper motion and OC's average proper motion is within 3$\sigma$; (3) OC's age fulfills $\log (t \mbox{ yr}^{-1})<8.6$; (4) the parallax of Cepheid is within the parallax range of OC members; (5) Cepheid is located in the blue-loop stage of the cluster. After selection, we obtained 33 high-probability OC Cepheids, 13 of which are newly discovered. Among these Cepheids, CM Sct is a first-crossing Cepheids with strict constraints of OC's membership and age. DP Vel is likely a fourth- or fifth-crossing Cepheids, but further evidence is needed to confirm it.

We obtained the distance, extinction, and age of OCs based on the isochrone fitting method. The age errors are $0.1-0.3$ dex and the distance error is around 0.1 mag. Based on OC's distances, we determined the mid-infrared PLR, $\langle M_{W1} \rangle = -(3.274 \pm 0.090) \log P-(-2.567 \pm 0.080), \sigma=0.075$, and the {\sl Gaia}-band PWR. The two relations generally agree with the previous works. The zero-point difference of $W1$-band PLR is around $-0.063$ mag and need a better study with future's database.  {\sl Gaia}-band PWR has a little larger scatter comparing to infrared PLRs. Based on OC's age, we obtained a direct PAR $\log t  =  -(0.638 \pm 0.063) \log P+(8.569 \pm 0.057), \sigma=0.073$ for fundamental Cepheids. The first-overtone Cepheid is 32\% (0.14 dex) younger than the fundamental Cepheid at a given period. Using Cepheid's PAR, the formation and evolution of the Milky Way's disc and spiral arms can be constrained by thousands of Cepheids.

The unprecedented accuracy of $\sl Gaia$ data provides an opportunity to well constrain both the membership and parameters of OC at the same time. With hundreds to thousands of new OCs in the future, we look forward to discovering dozens of Cepheids with age information. The zero point of PAR will be better constrained, and the knowledge of Cepheids' evolution state will be more complete.

\section*{acknowledgments} We thank the anonymous referee for comments that helped us
improve the paper. We are grateful for research support from the National
  Natural Science Foundation of China through grants 11903045, 11633005. This work is also supported by Major Science and Technology Project of Qinghai Province 2019-ZJ-A10.

This work has made use of data from the European Space Agency (ESA) mission Gaia (https://www.cosmos.esa.int/gaia), processed by the Gaia Data Processing and Analysis Consortium (DPAC, https://www.cosmos.esa.int/web/gaia/dpac/consortium). Funding for the DPAC has been provided by national institutions, in particular the institutions participating in the Gaia Multilateral Agreement.

\section*{Data availability}
The Gaia EDR3 data is available at https://gea.esac.esa.int/archive/. The data underlying this article will be shared on reasonable request to the corresponding author.

\appendix 
\section{Table for Cepheids that are not assumed as OC Cepheids in this work.}
\onecolumn
\clearpage
\begin{landscape}
\begin{table}
\footnotesize
\caption{Unlikely and possible OC Cepheids.\label{c2table3}}
\begin{tabular}{@{}lcccccccc@{}}
  \hline
Cluster  &  $G_{\rm obs}-G_{\rm theo}$   & $(BP-RP)_{\rm obs}-(BP-RP)_{\rm theo}$ & p: possible OC Cepheid   &Cepheid&  Period  & Seperation  & $\Delta \mu$ & $\Delta parallax$  \\
         &  mag                          &mag                                     & b: background Cepheid    &&  day     &   arcmin    &              &                    \\
         &                               &                                        & n: not like Cepheid      &&          &             &              &                    \\
         &                               &                                        & o: OC with bad parameters&&          &             &              &                    \\  
  \hline
BH 150       &1 &   &p&OGLE GD1267.01.00029        &1.36662 &2.2 &1.0&0.3\\  
FSR 0172     &  &1.3&p&StRS 329                    &18.31543&1.7 &1.1&1.7\\            
Feibelman 1  &  &0.5&p&2MASS J20173903+3804143     &5.05214 &2.4 &2.0&2.0\\            
UBC 293      &$-1$&   &p&V659 Cen                    &5.62325 &12.6&0.9&1.2\\
UBC 600      &2 &   &p&V1100 Cas                   &4.70346 &28.7&0.7&1.9\\         
Hogg 17      &8 &2  &b&OGLE GD-CEP-991             &12.71923&6.1 &0.8&1.0\\            
Kronberger 81&8 &   &b&TYC 3966-454-1              &2.52310 &3.6 &1.8&0.8\\        
NGC 4052     &7 &3  &b&OGLE GD-CEP-740             &8.28794 &4.0 &1.3&1.0\\            
NGC 6631     &4 &4  &b&OGLE BLG-CEP-164            &13.62695&3.9 &0.7&0.0\\            
Trumpler 22  &$>8$&$>4$ &b&TYC 9006-1244-1             &5.65178 &16.4&0.5&0.4\\        
UBC 266      &12&   &b&[G93c] 19                   &0.48045 &7.1 &0.5&0.5\\        
UBC 608      &4 &   &b&TYC 3722-6-1                &1.82183 &1.4 &1.2&0.8\\        
UBC 667      &9 &4  &b&OGLE GD-CEP-1157            &13.38194&4.0 &1.2&0.8\\            
UBC 94       &11&   &b&OGLE BLG-CEP-131            &10.21399&11.9&2.5&1.0\\ 
Alessi 21    &12&   &n&OGLE GD-CEP-95              &0.32751 &50.6&1.0&1.5\\            
BDSB96       &8 &   &n&OGLE GD1693.03.00051        &3.92038 &1.6 &0.5&0.4\\           
BH 121       &8 &   &n&Gaia DR2 5333575119290099200&2.06106 &8.4 &1.2&0.4\\           
Collinder 95 &12&   &n&OGLE GD-CEP-42              &1.51619 &10.9&1.3&0.8\\           
Czernik 41   &4 &   &n&Gaia DR2 2026716799817156352&2.93868 &2.3 &1.9&2.0\\           
Harvard 16   &7 &   &n&OGLE BLG-CEP-41             &0.23961 &6.7 &2.4&1.0\\           
IC 2395      &12&   &n&OGLE GD2101.28.00108        &1.61966 &24.9&0.2&0.5\\           
IC 2581      &4 &   &n&V348 Car                    &2.78109 &4.9 &0.4&0.9\\           
IC 2948      &10&   &n&Gaia DR2 5333575119290099200&2.06106 &3.4 &2.3&0.5\\       
NGC 3603     &7 &   &n&OGLE GD-CEP-641             &0.35879 &4.8 &0.5&1.5\\           
NGC 6193     &10&1  &n&OGLE GD1108.13.11919        &1.98486 &6.6 &0.4&1.0\\           
NGC 6322     &12&   &n&OGLE BLG-CEP-109            &1.67235 &1.7 &2.5&2.3\\           
Trumpler 14  &9 &   &n&Trumpler 14 VBF 452         &4.57045 &1.3 &0.3&0.1\\    
UBC 553      &13&   &n&OGLE GD-CEP-1194            &1.92093 &4.0 &0.4&1.6\\           
UBC 587      &12&1  &n&TYC 3178-1095-1             &3.05869 &16.5&0.9&0.3\\ 
Schuster 1   &  &   &o&OGLE GD-CEP-436             &4.05216 &3.2 &0.4&1.3\\
\hline
\end{tabular}
\end{table}
\end{landscape}
\section{Graphical representations of diagnostic diagrams for other
  open cluster Cepheids.}

\clearpage
\begin{figure}
\includegraphics[width=160mm]{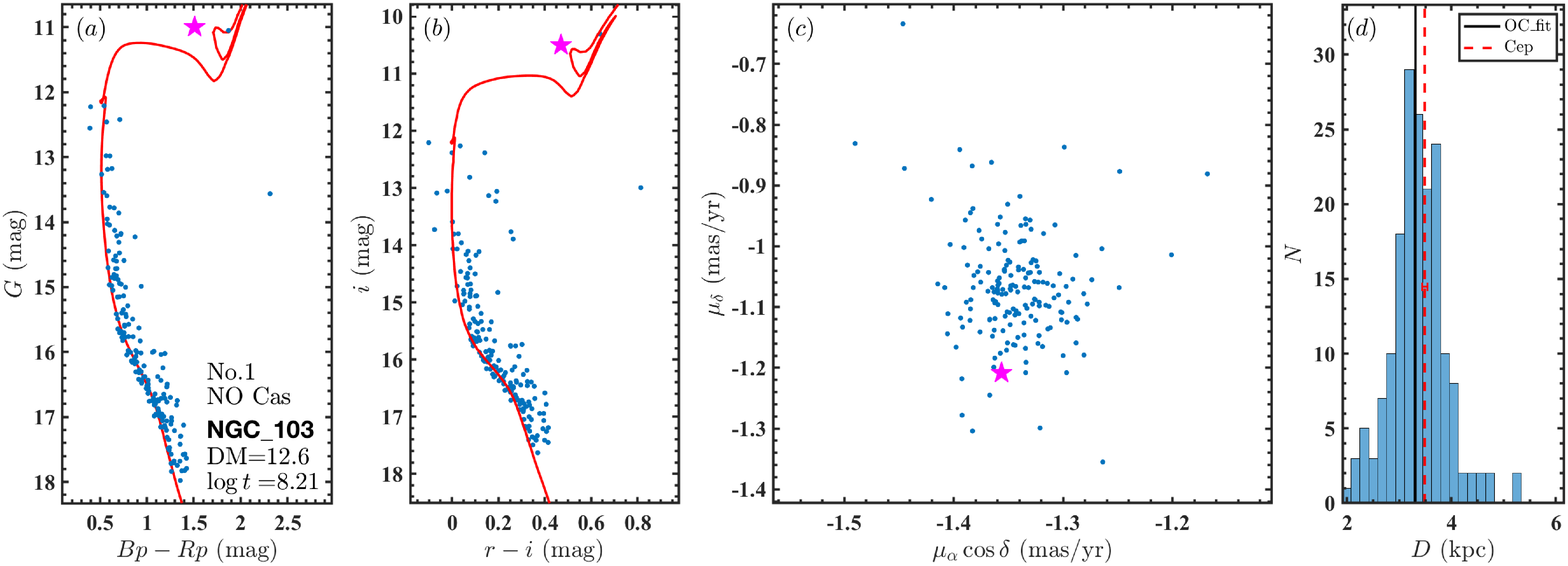}
\includegraphics[width=160mm]{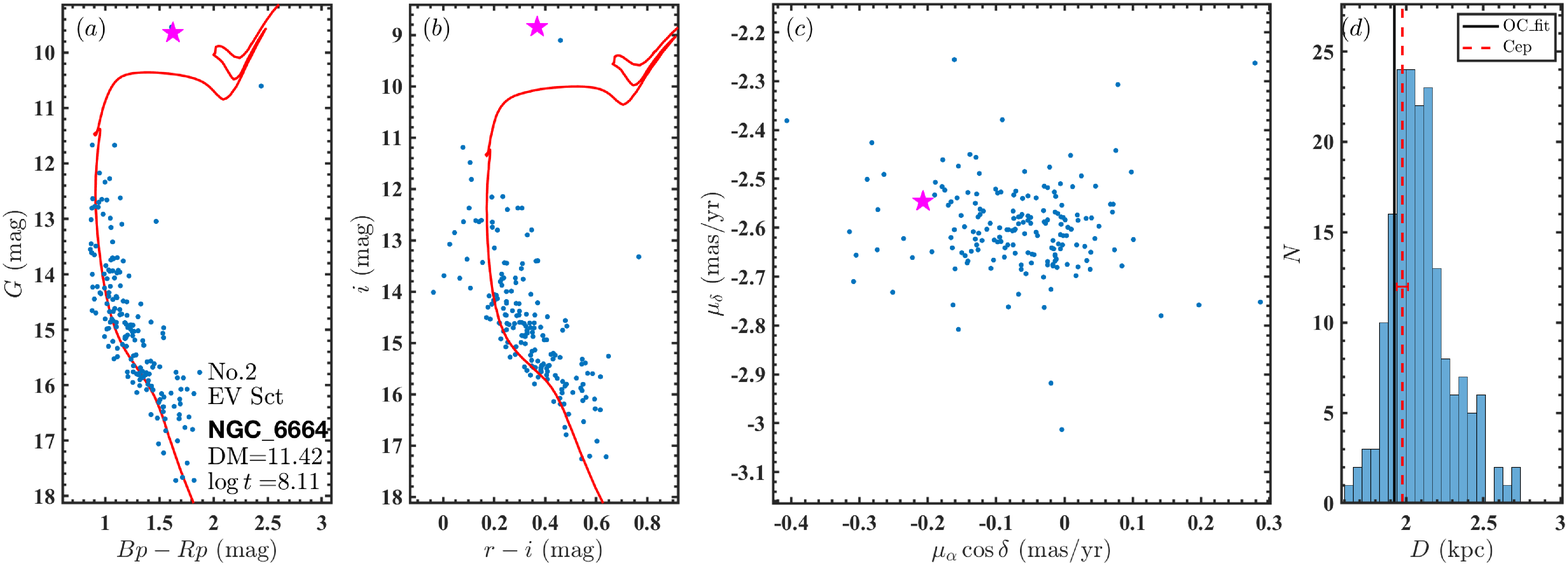}
\includegraphics[width=160mm]{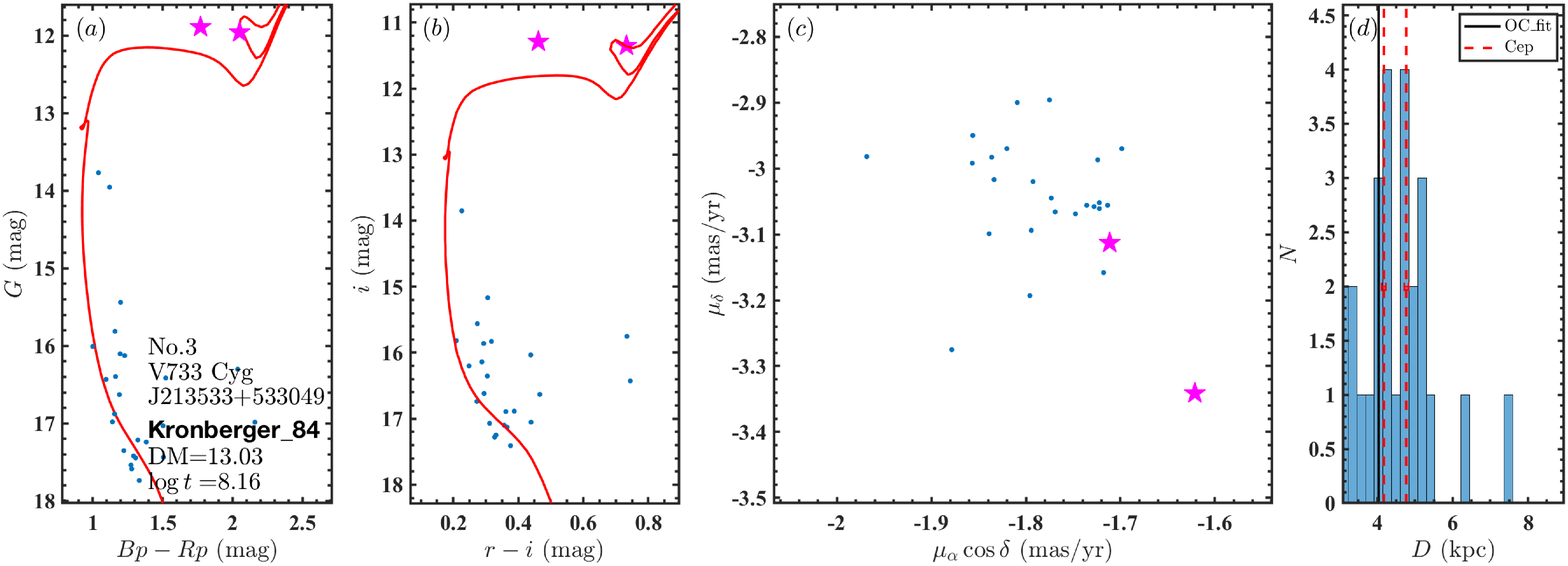}
\includegraphics[width=160mm]{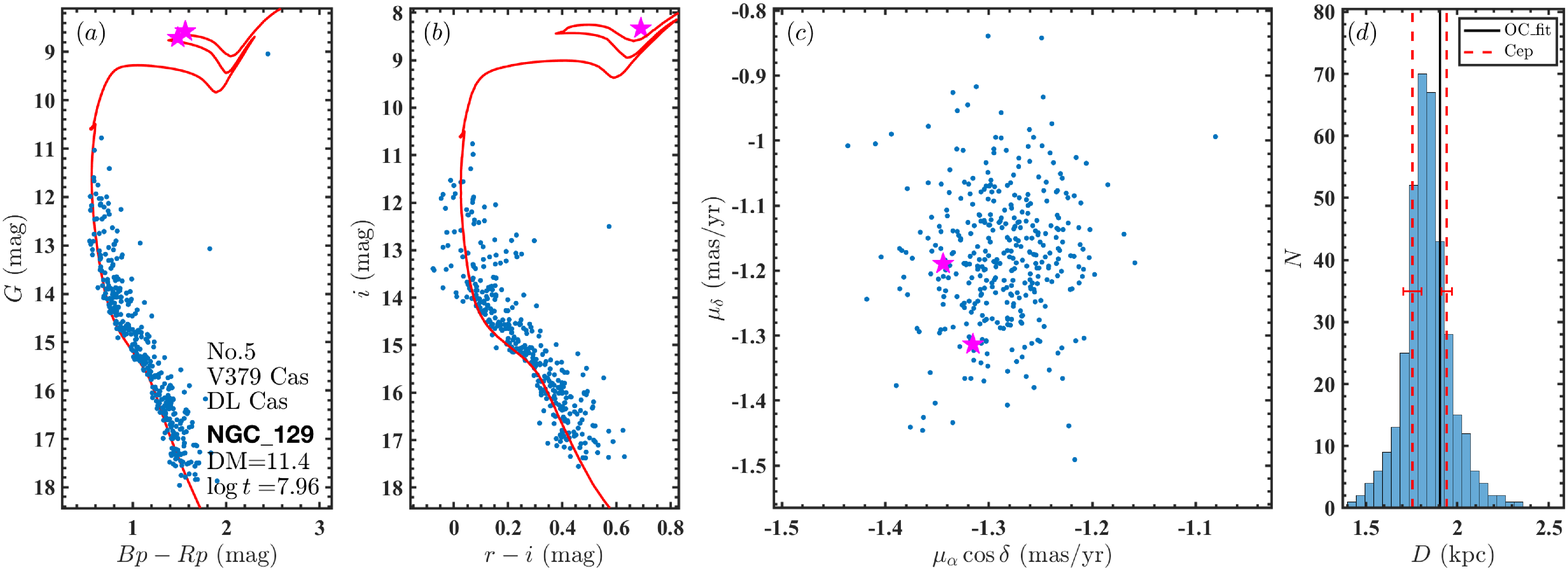}
\caption{Diagnostic diagrams for Cepheids in OC NGC 103, NGC 6664, Kronberger 84, and NGC 129.}
\end{figure}

\clearpage

\begin{figure}
\includegraphics[width=160mm]{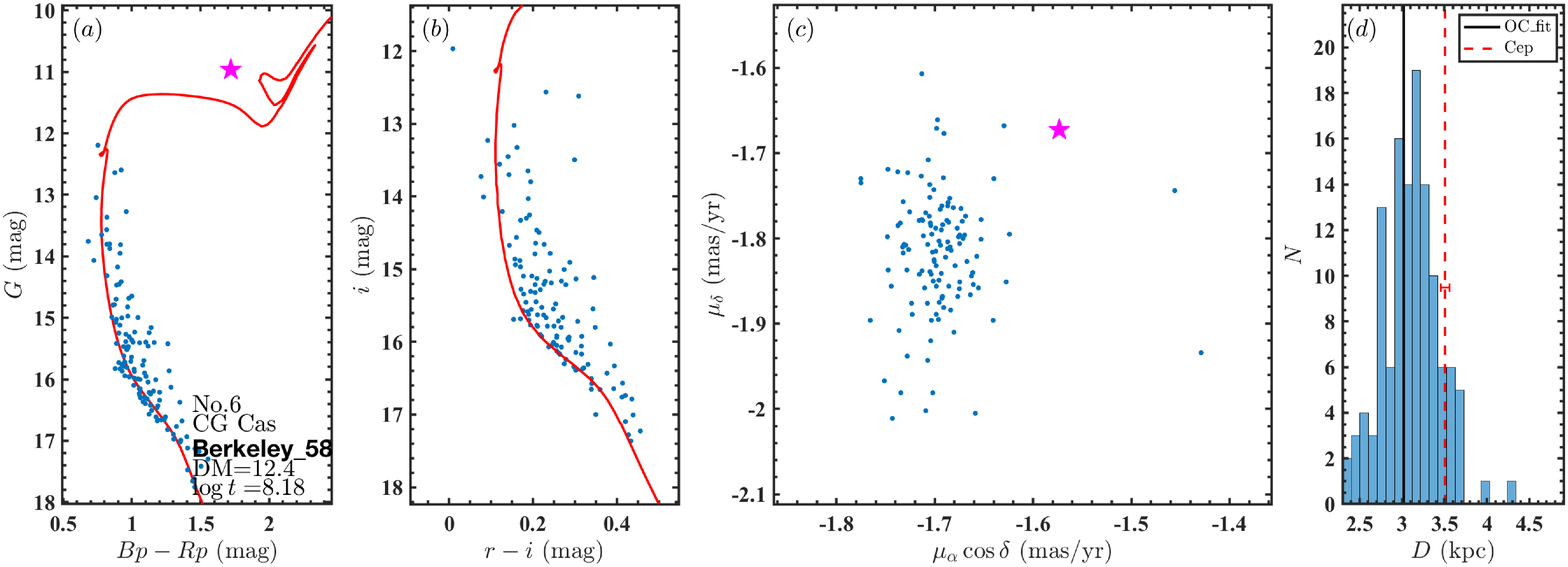}
\includegraphics[width=160mm]{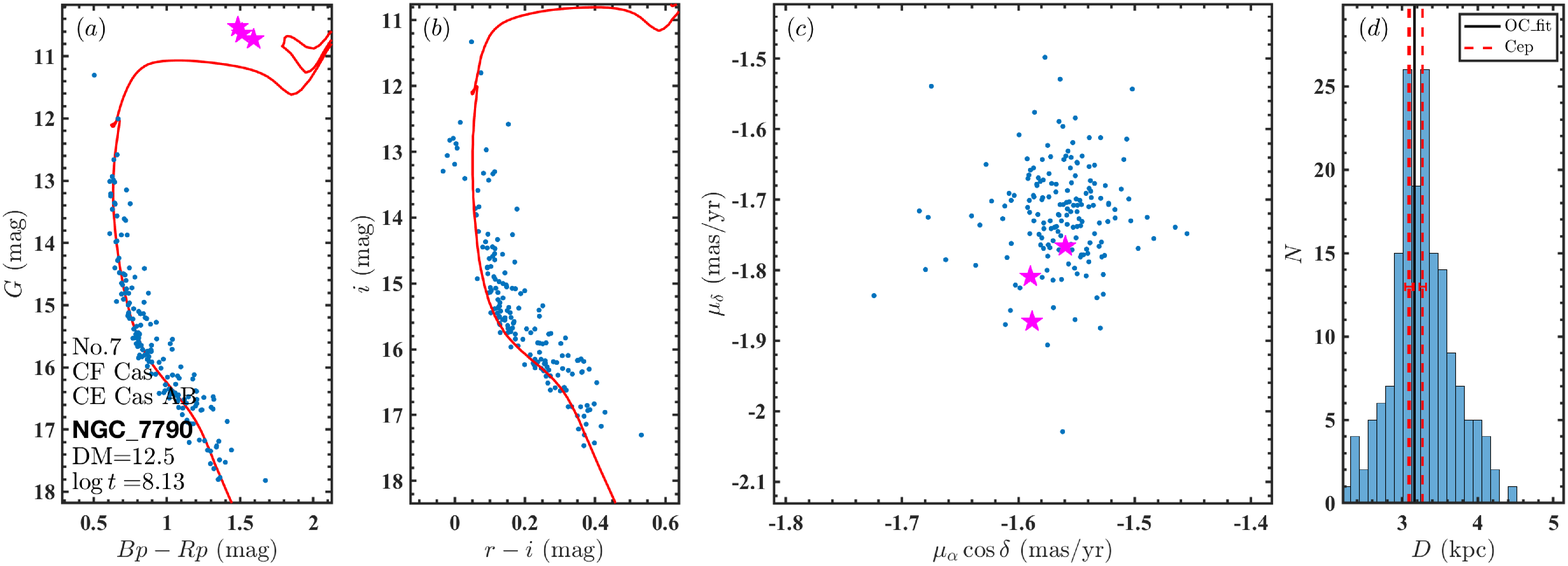}
\includegraphics[width=160mm]{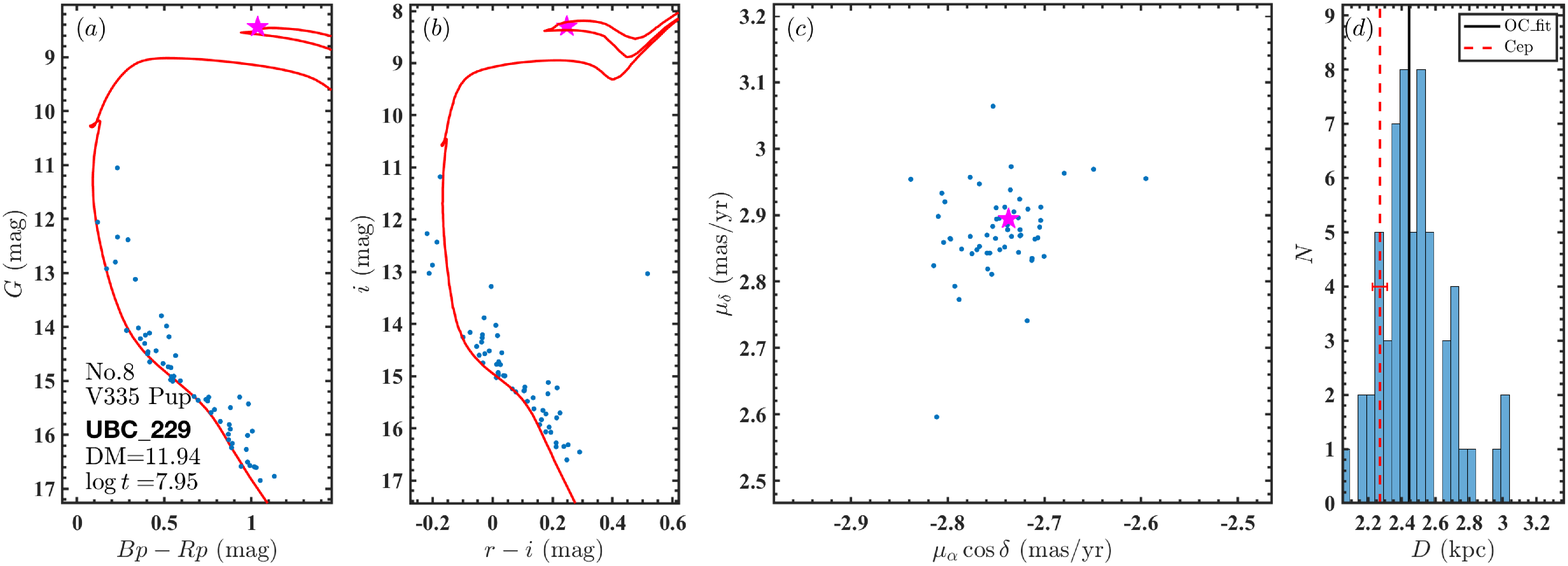}
\includegraphics[width=160mm]{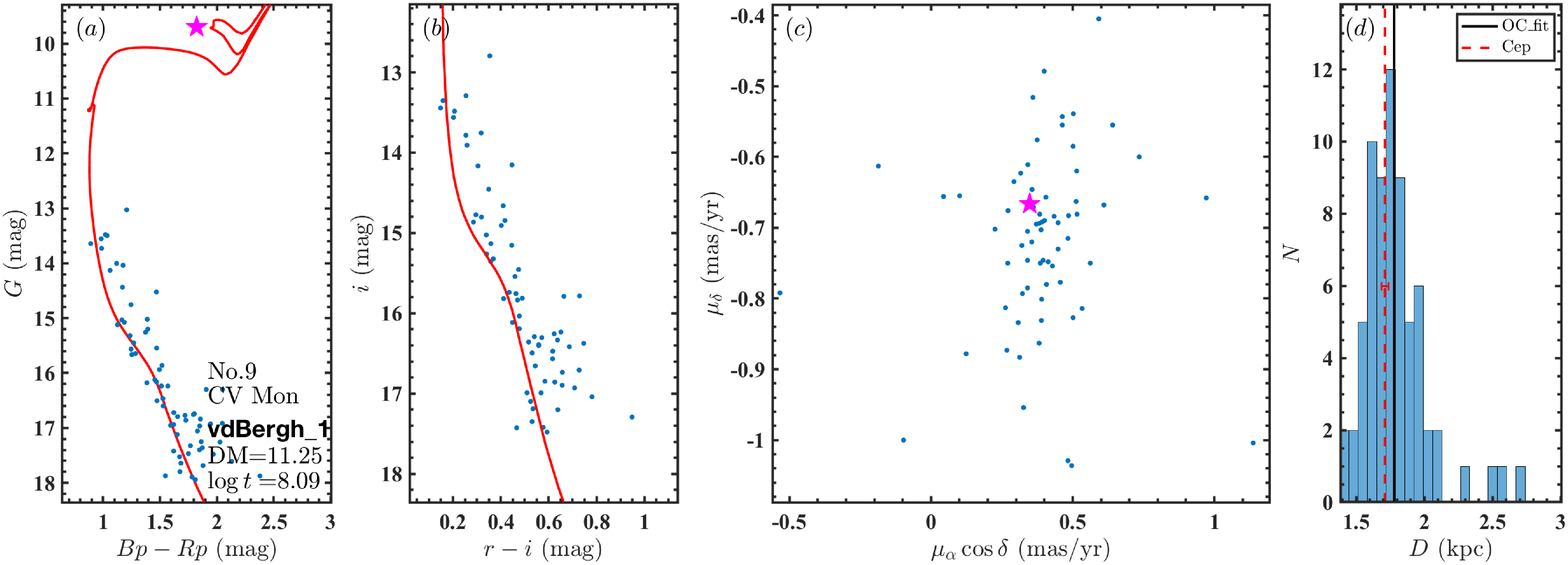}
\caption{Diagnostic diagrams for Cepheids in OC Berkeley 58, NGC 7790, UBC 229, and vdBergh 1.}
\end{figure}

\clearpage

\begin{figure}
\includegraphics[width=160mm]{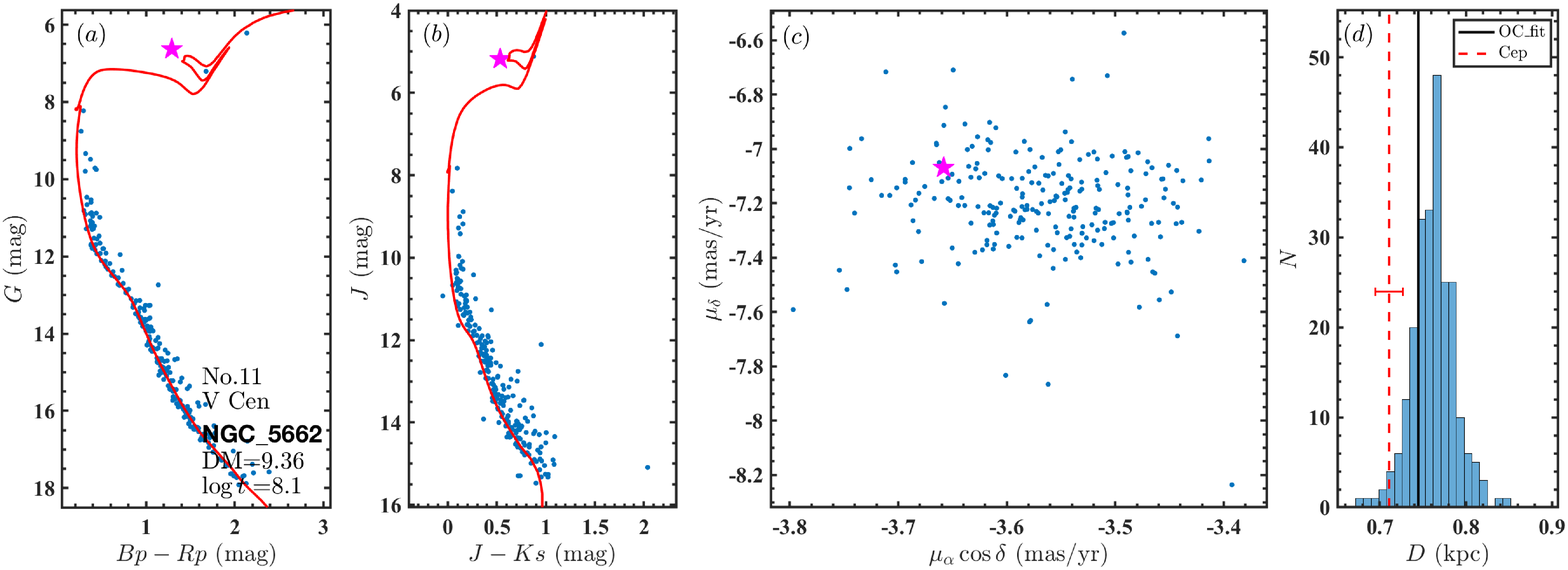}
\includegraphics[width=160mm]{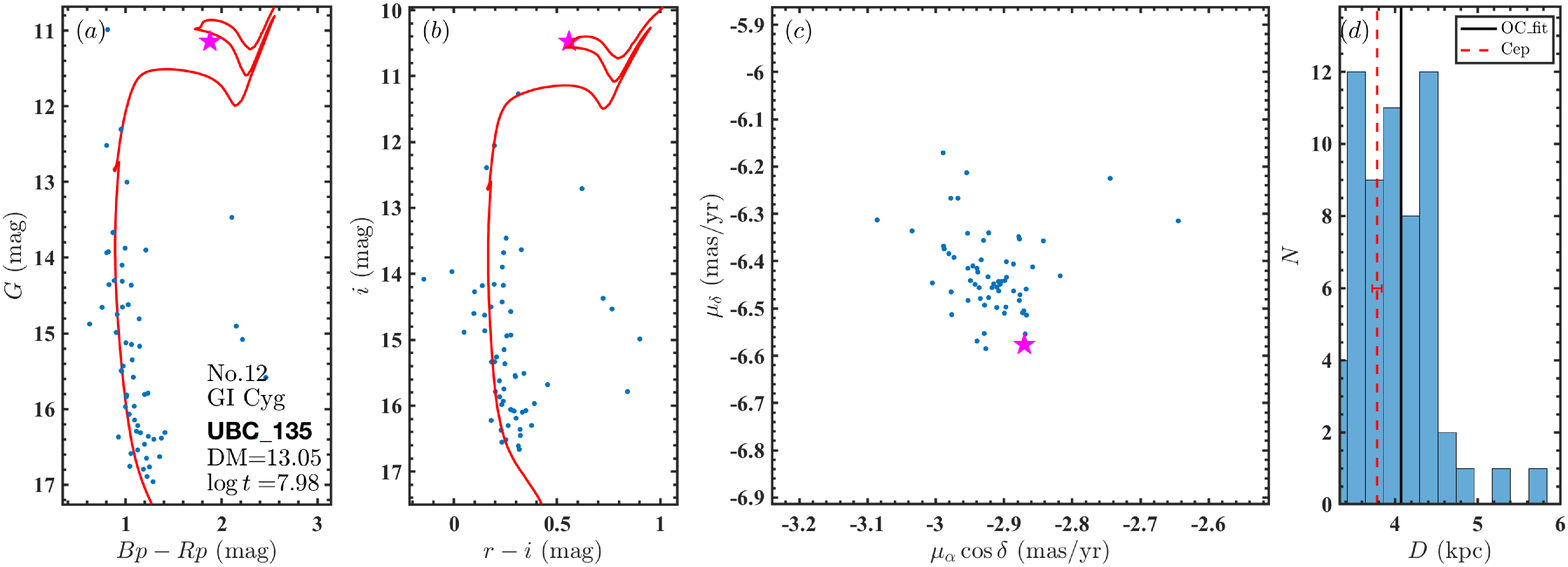}
\includegraphics[width=160mm]{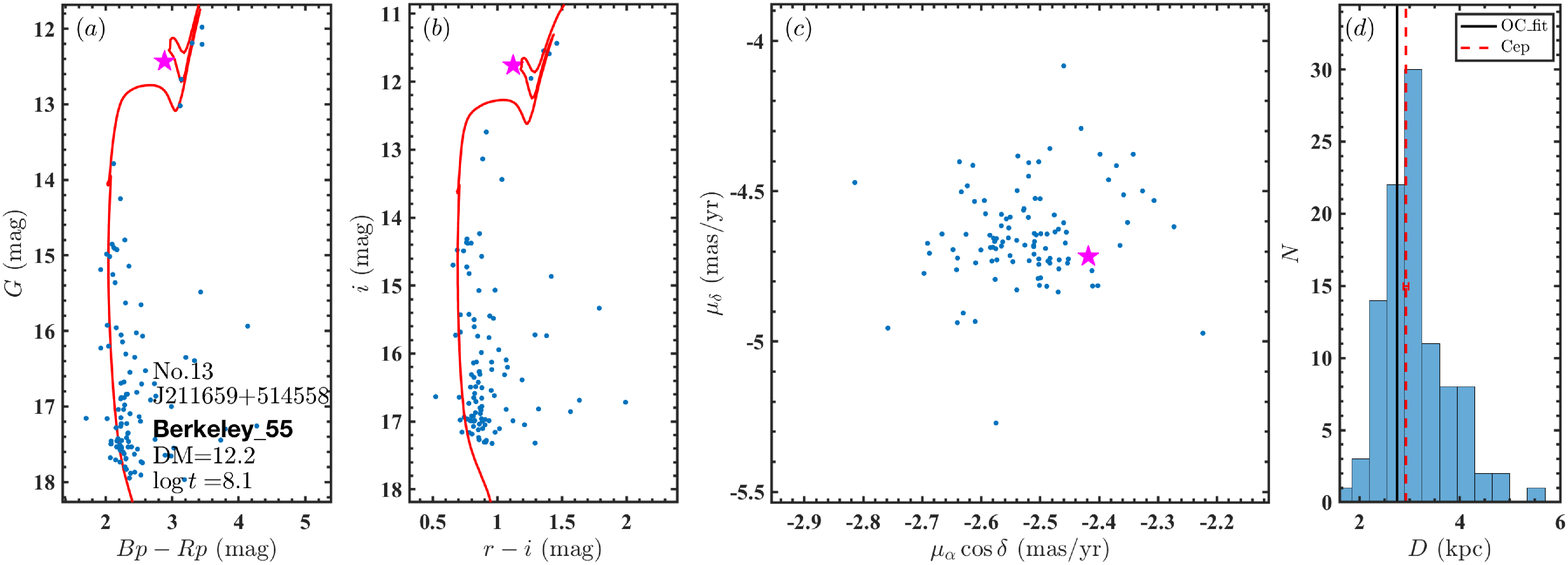}
\includegraphics[width=160mm]{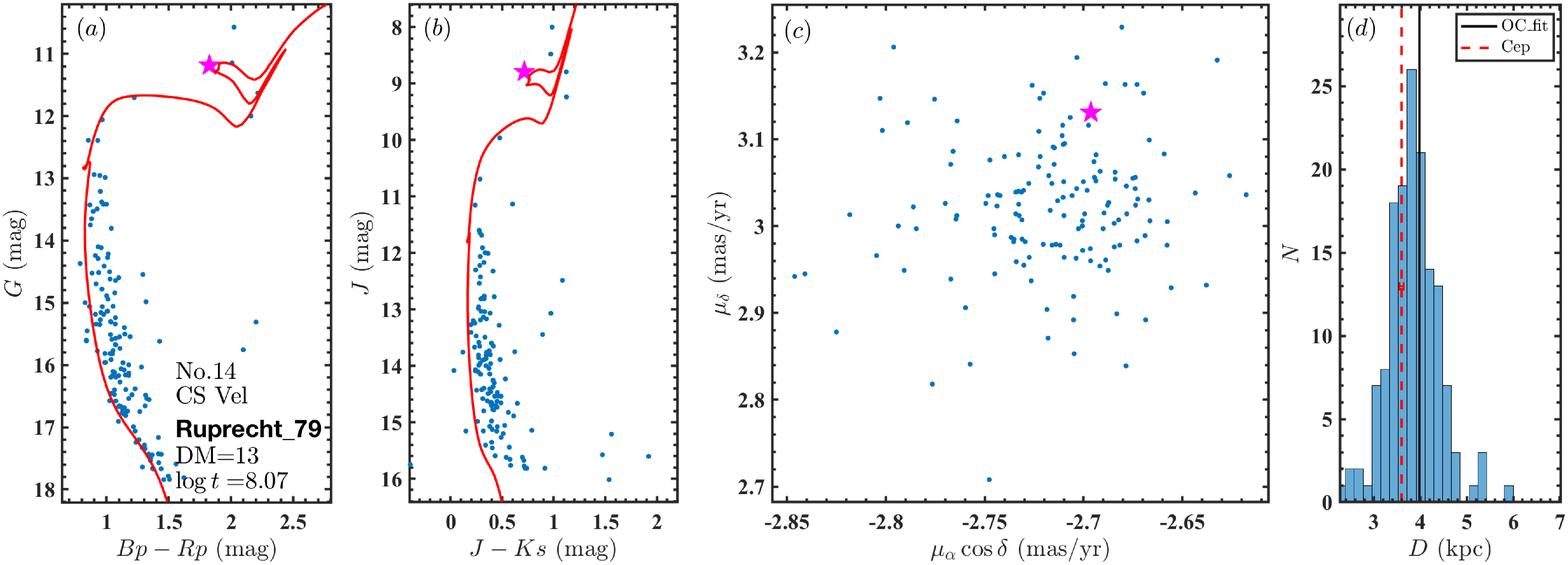}
\caption{Diagnostic diagrams for Cepheids in OC NGC 5662, UBC 135, Berkeley 55, and Ruprecht 79.}
\end{figure}

\clearpage

\begin{figure}
\includegraphics[width=160mm]{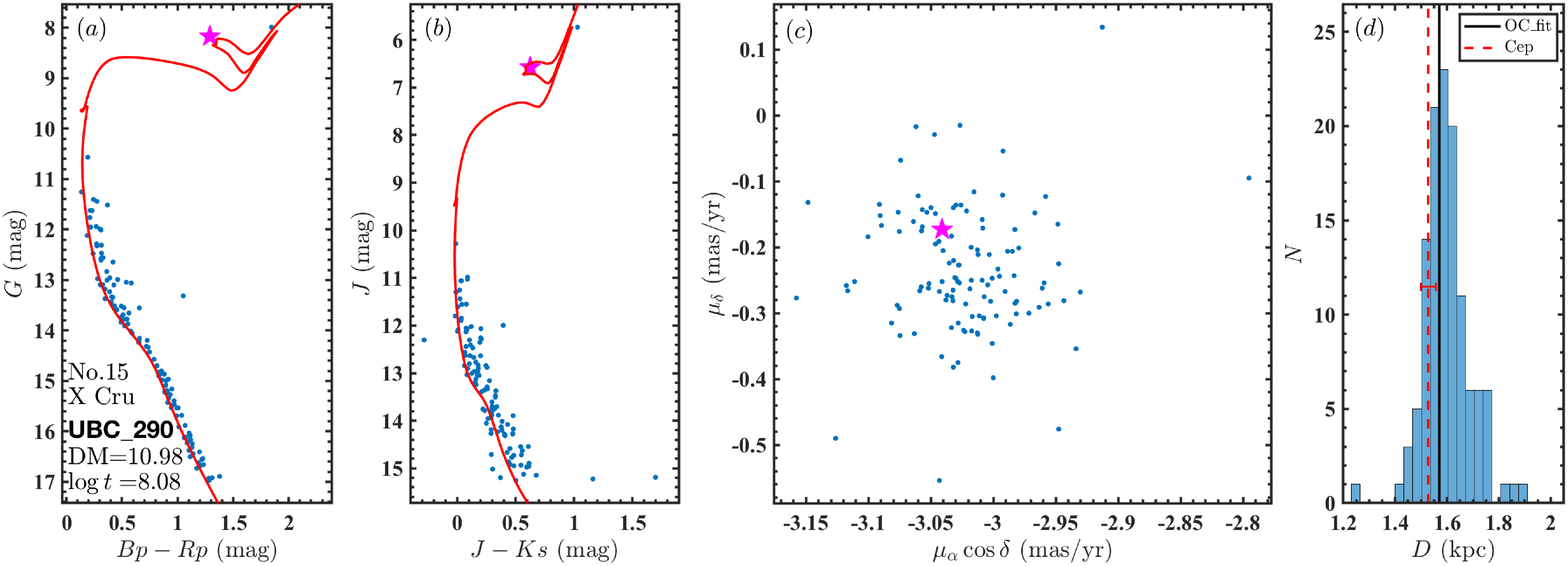}
\includegraphics[width=160mm]{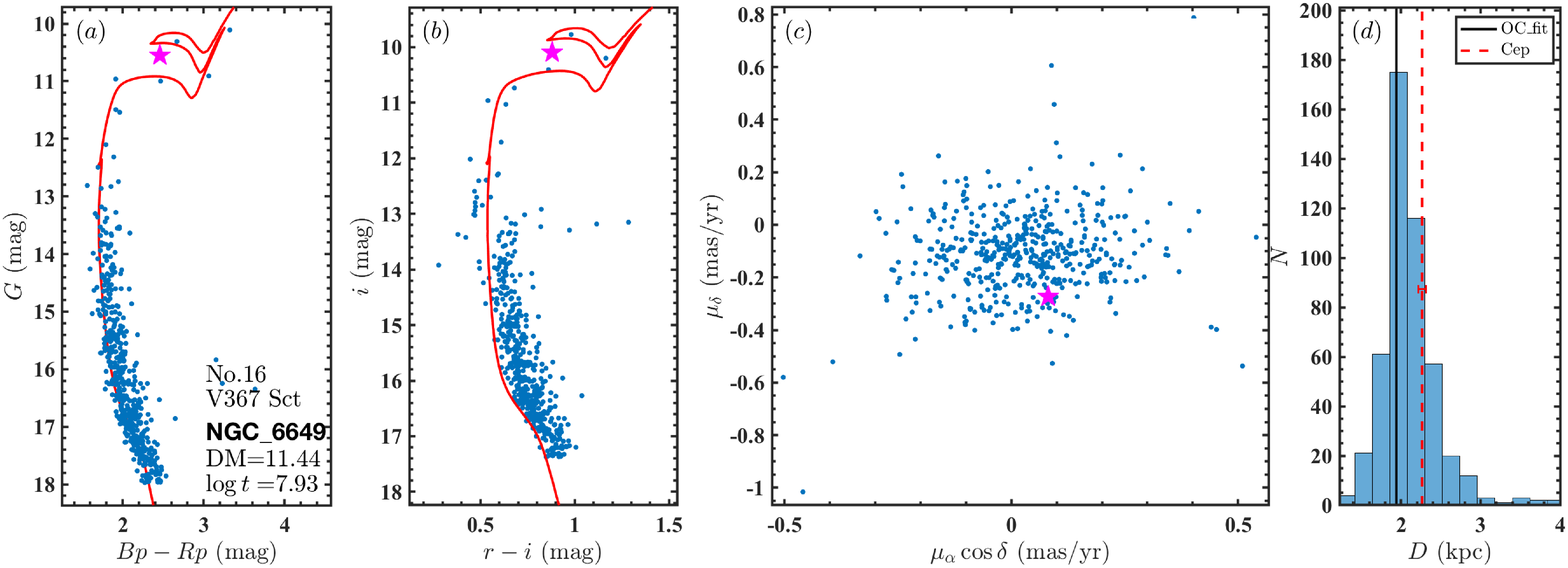}
\includegraphics[width=160mm]{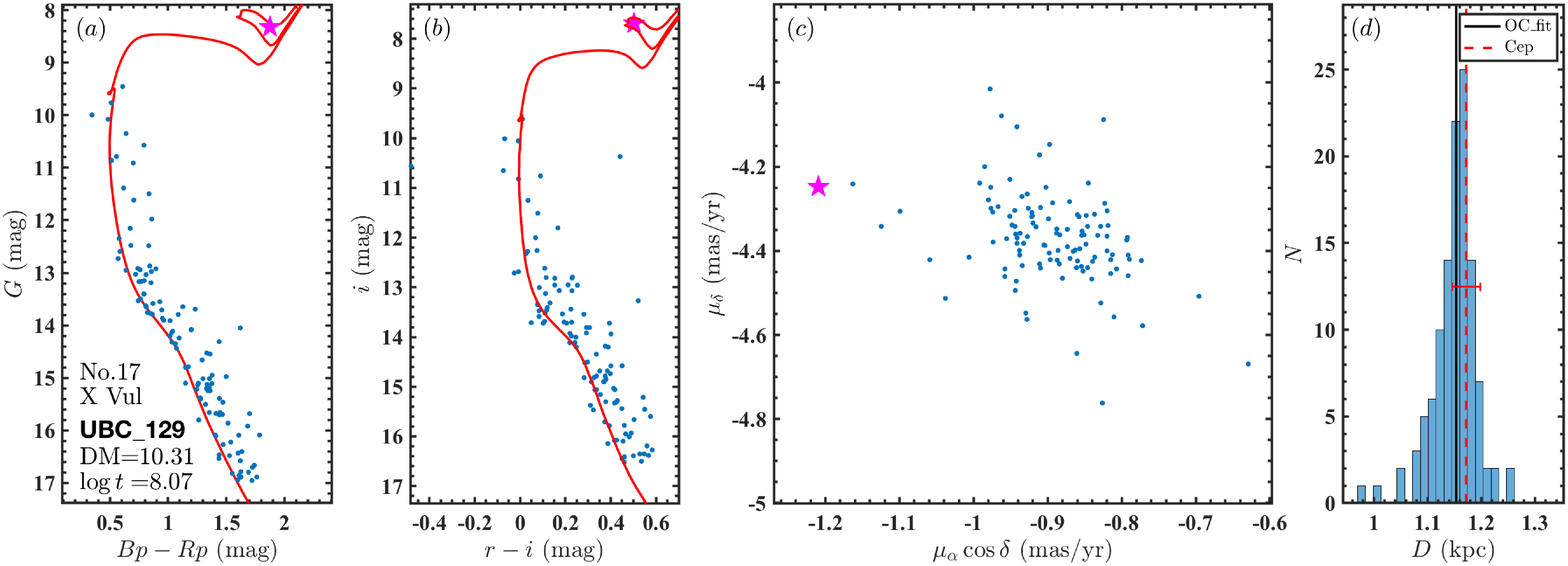}
\includegraphics[width=160mm]{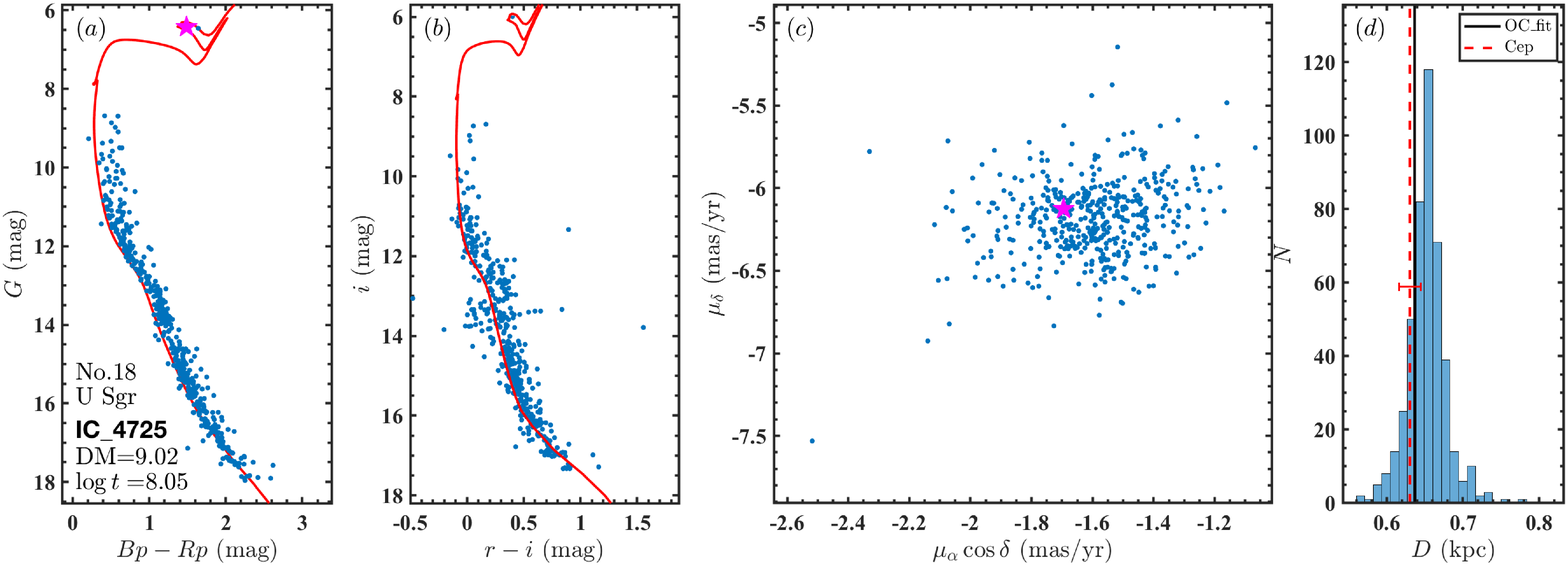}
\caption{Diagnostic diagrams for Cepheids in OC UBC 290, NGC 6649, UBC 129, and IC 4725.}
\end{figure}

\clearpage

\begin{figure}
\includegraphics[width=160mm]{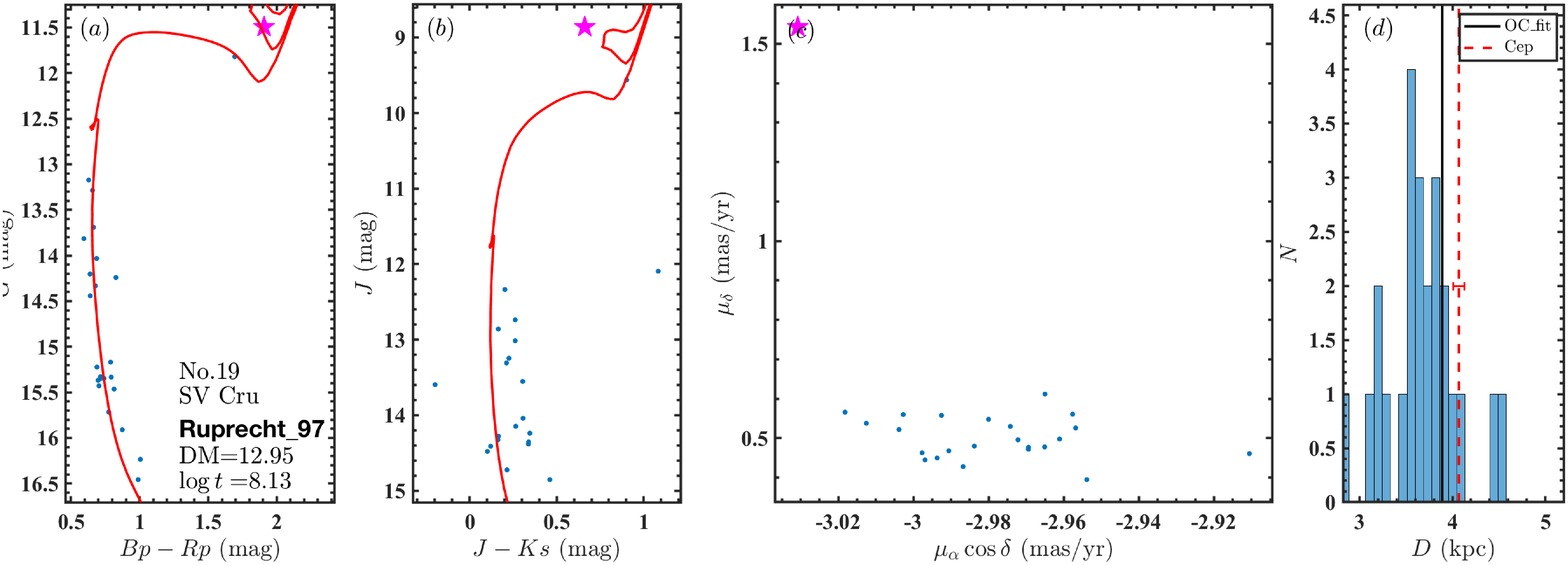}
\includegraphics[width=160mm]{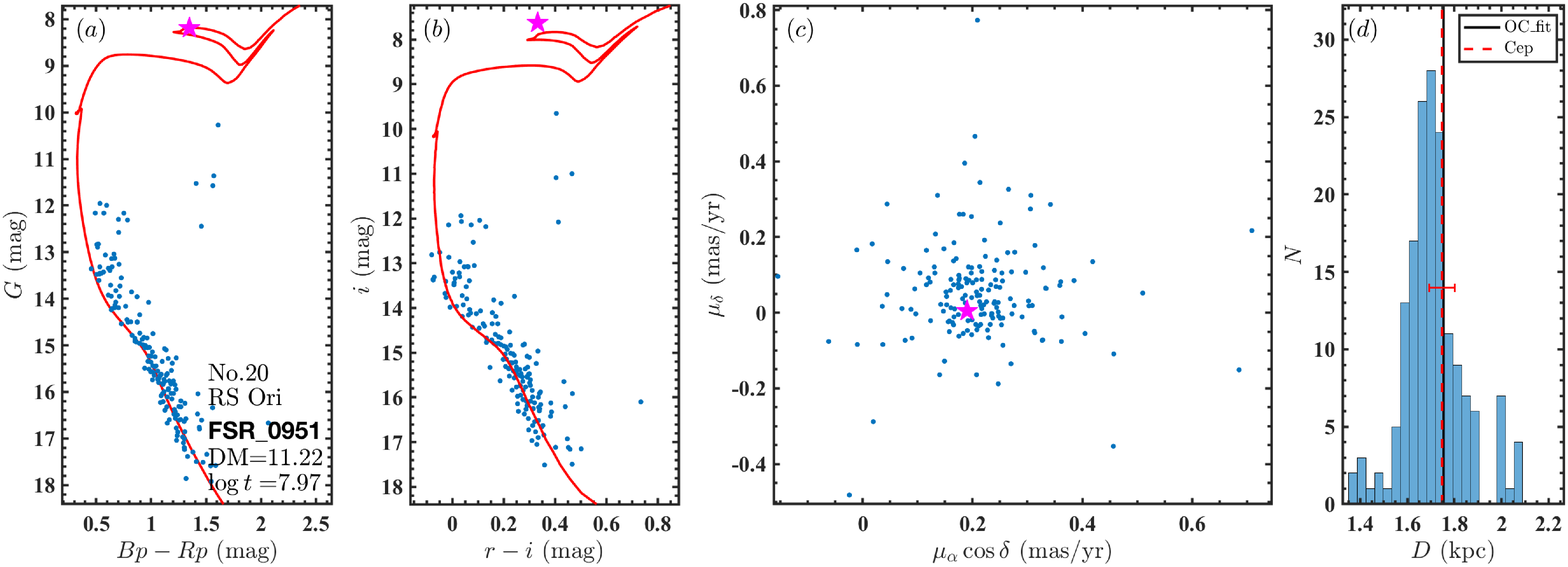}
\includegraphics[width=160mm]{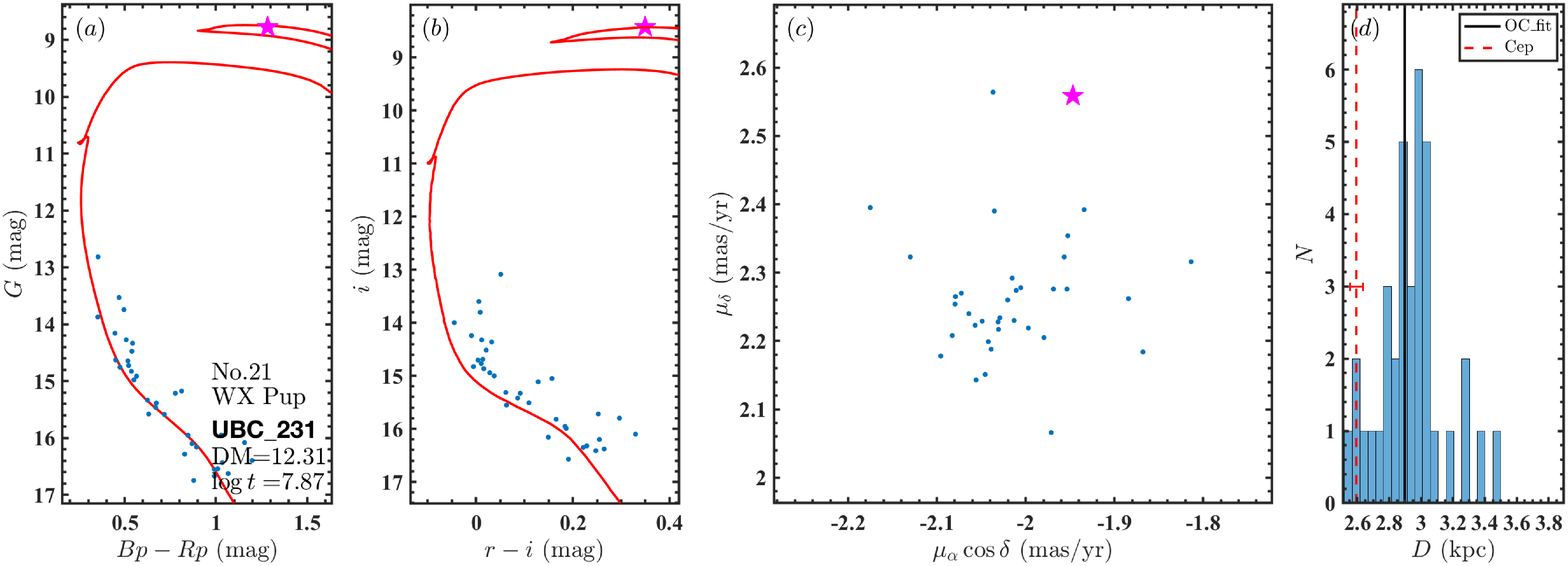}
\includegraphics[width=160mm]{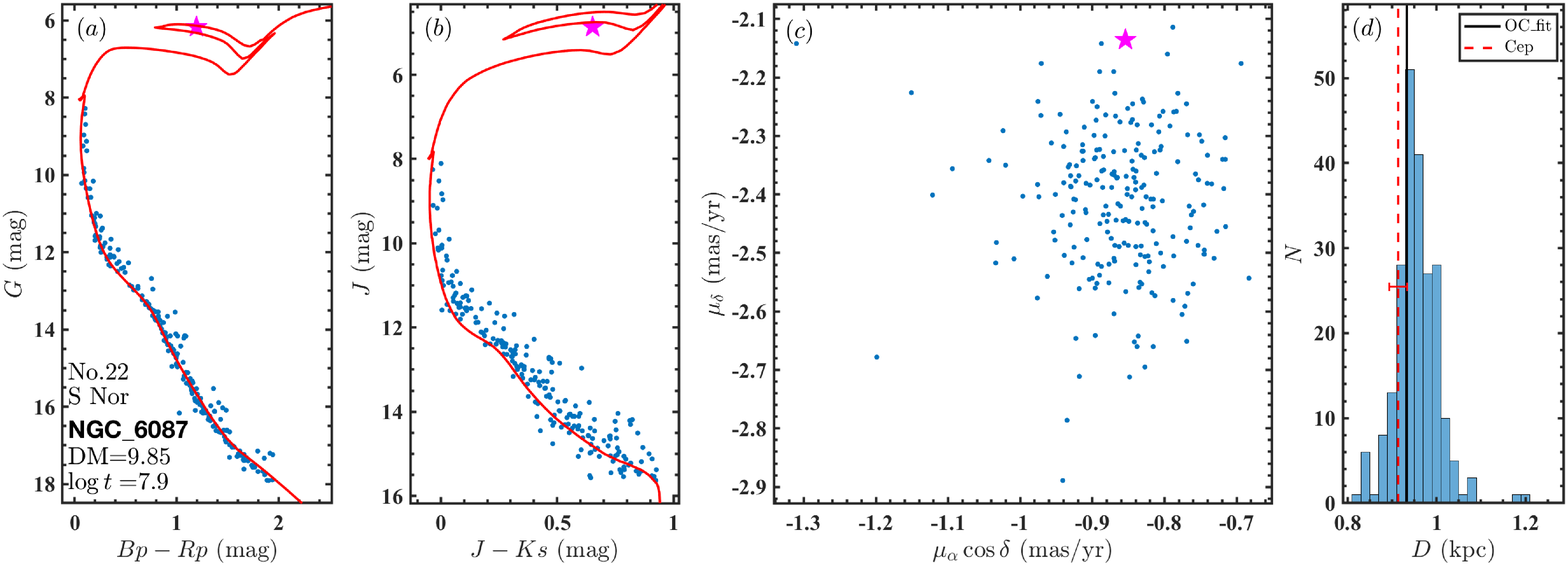}
\caption{Diagnostic diagrams for Cepheids in OC Ruprecht 97, FSR 0951, UBC 231, and NGC 6087.}
\end{figure}

\clearpage

\begin{figure}
\includegraphics[width=160mm]{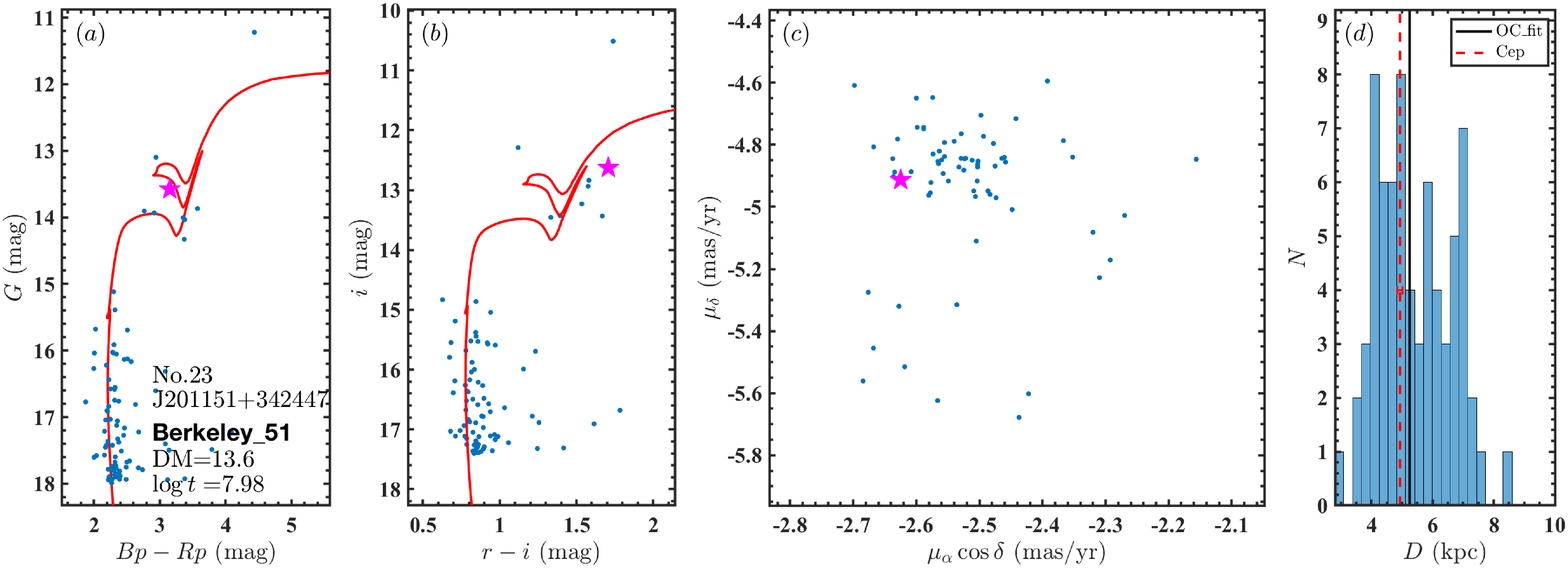}
\includegraphics[width=160mm]{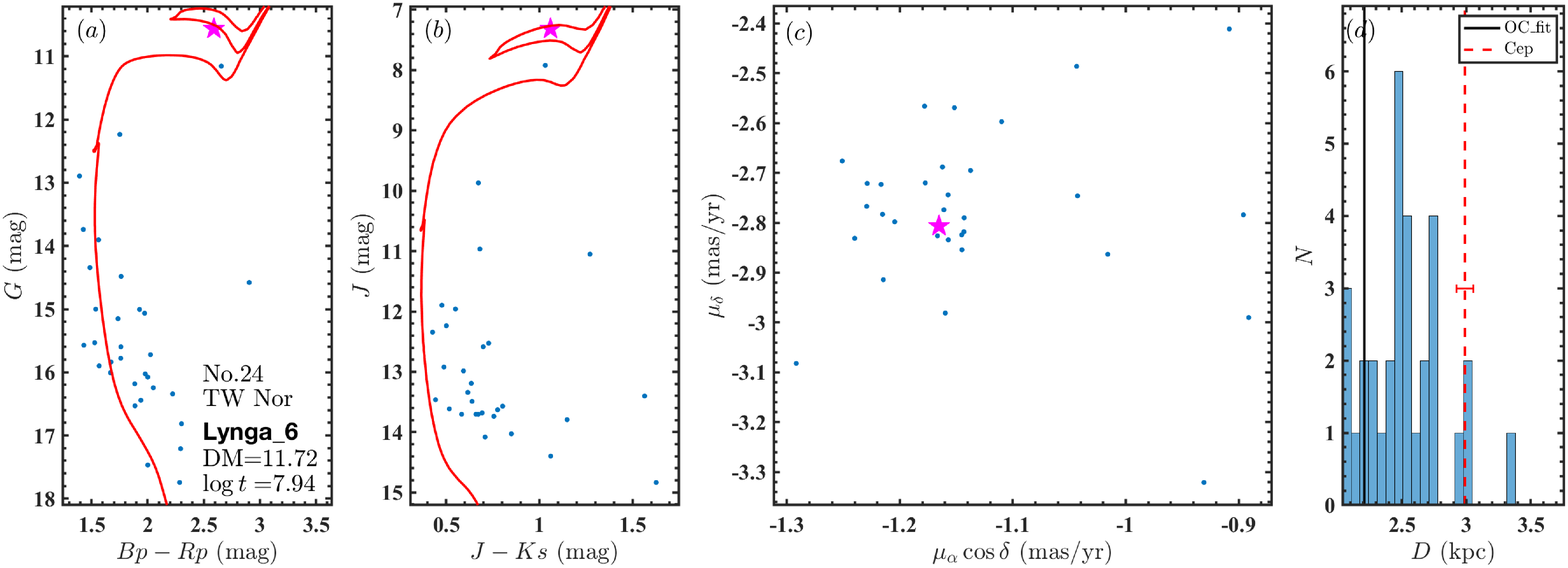}
\includegraphics[width=160mm]{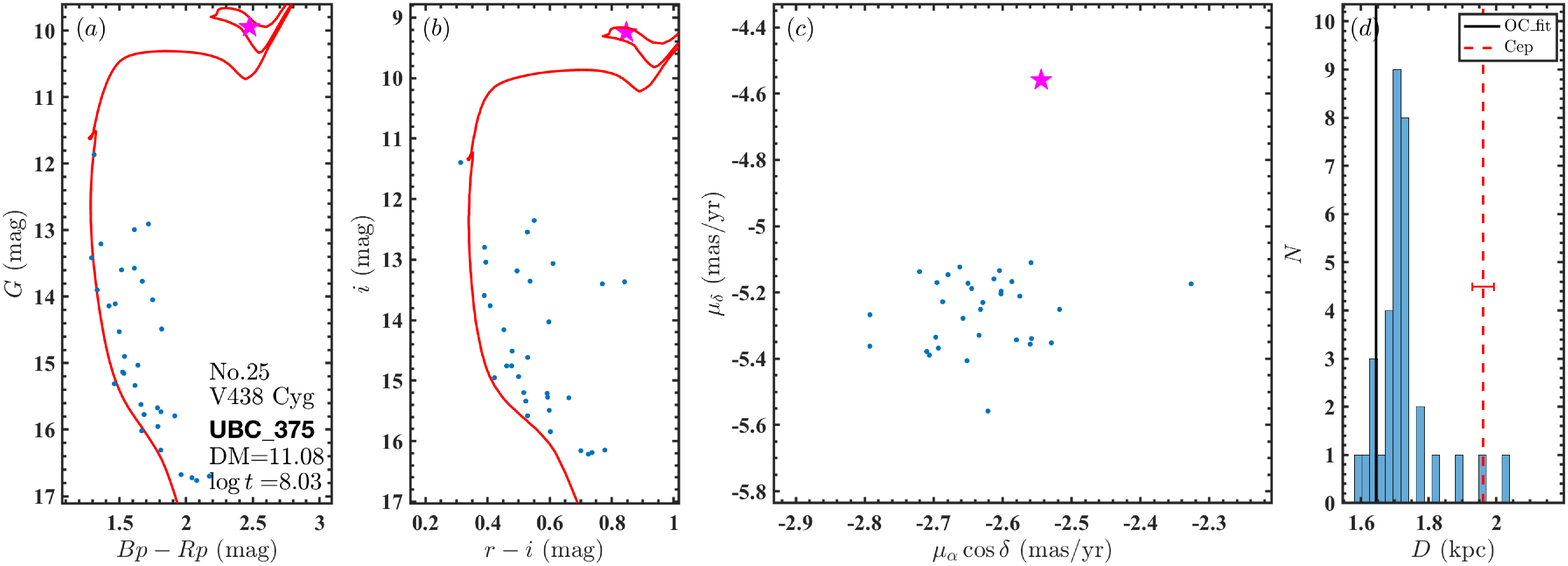}
\includegraphics[width=160mm]{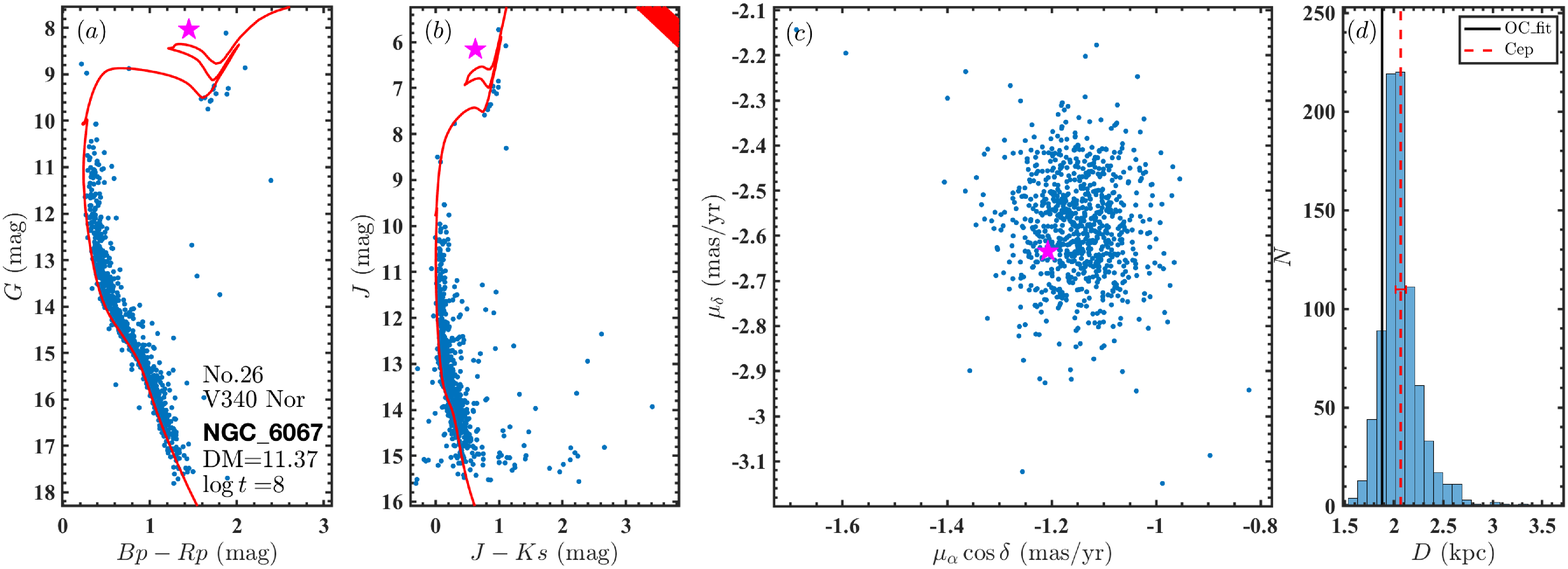}
\caption{Diagnostic diagrams for Cepheids in OC Berkeley 51, Lynga 6, UBC 375, and NGC 6067.}
\end{figure}

\clearpage

\begin{figure}
\includegraphics[width=160mm]{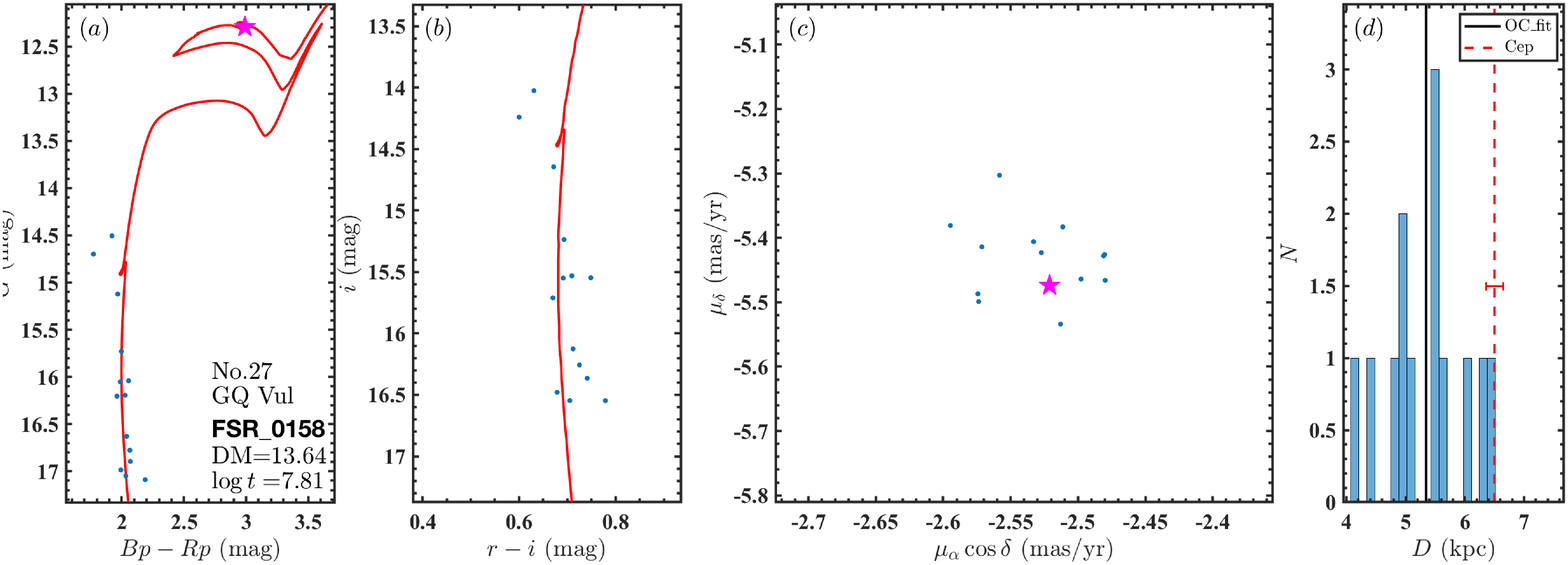}
\includegraphics[width=160mm]{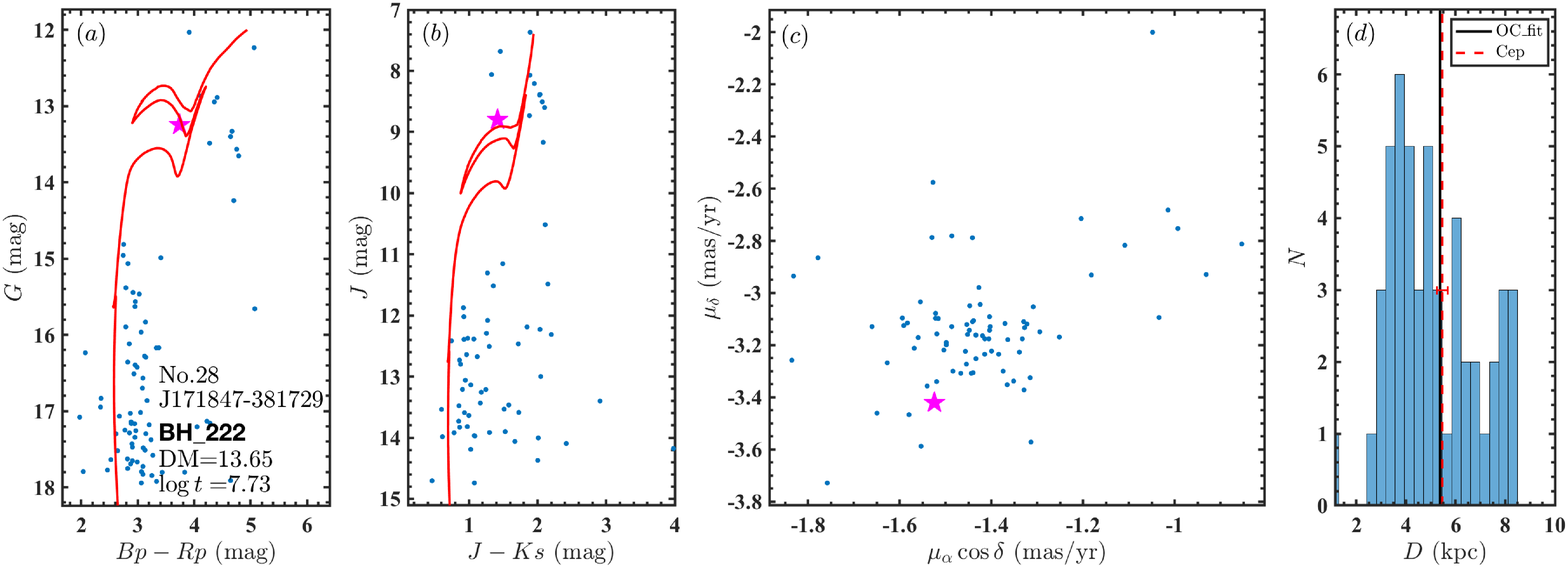}
\includegraphics[width=160mm]{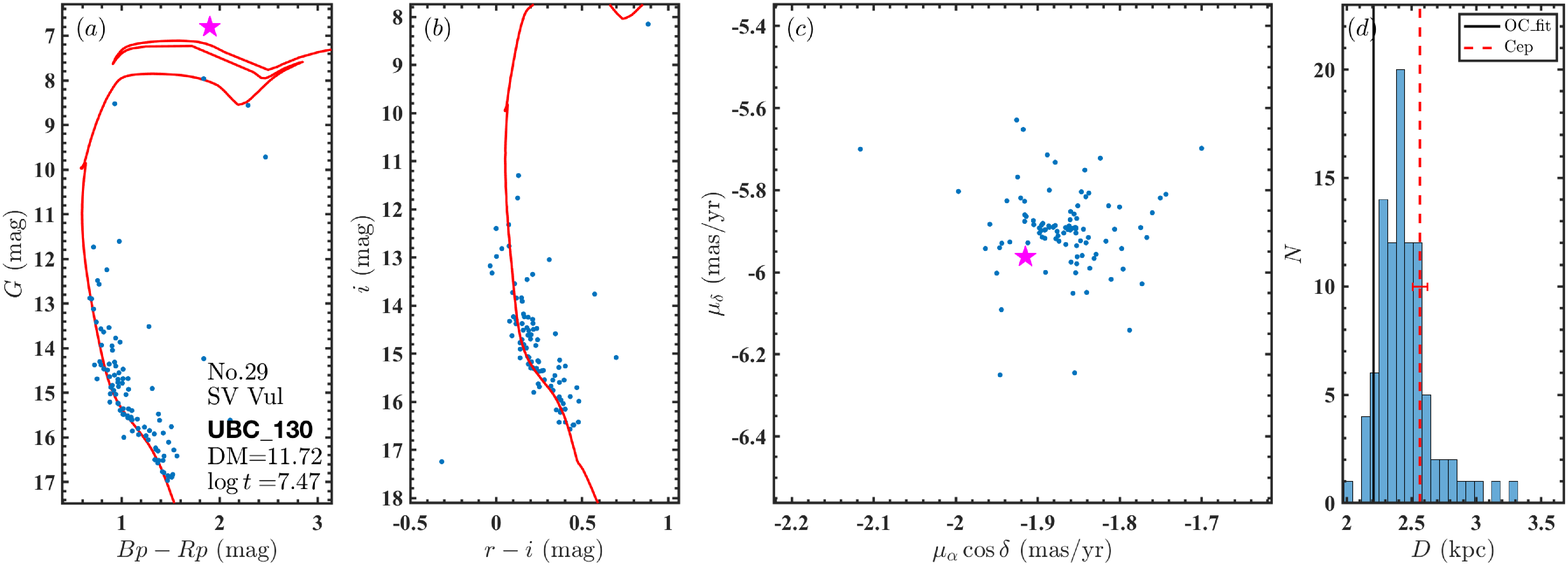}
\caption{Diagnostic diagrams for Cepheids in the OC FSR 0158, BH 222, UBC 130.}
\end{figure}

\end{document}